\documentclass[twocolumn,numberedappendix]{emulateapj}
\usepackage{apjfonts}

\usepackage{graphicx}
\usepackage{mathrsfs}
\usepackage{natbib}
\bibpunct[,]{(}{)}{;}{a}{}{,}

\def\totd{{\mathrm{d}}}
\def\rs0{{r_{\rm s0}}}

\shorttitle{HYDRODYNAMICS OF CORE-COLLAPSE SUPERNOVAE}
\shortauthors{FERN\'ANDEZ}

\begin{document}

\title{Hydrodynamics of Core-Collapse Supernovae at the Transition to Explosion. \\I. Spherical Symmetry}
\author{Rodrigo Fern\'andez\altaffilmark{1}}
\affil{Institute for Advanced Study. Einstein Drive, Princeton, NJ 08540, USA.}
\altaffiltext{1}{Einstein Fellow}

\begin{abstract}
We study the transition to runaway expansion of an initially stalled core-collapse supernova
shock. The neutrino luminosity, mass accretion rate, and neutrinospheric radius are
all treated as free parameters. 
In spherical symmetry, this transition is mediated by a global non-adiabatic instability that develops on
the advection time and reaches non-linear amplitude. Here we perform high-resolution, time-dependent 
hydrodynamic simulations of stalled supernova shocks with realistic microphysics to analyze this transition.
We find that radial instability is a sufficient condition for runaway expansion if the
neutrinospheric parameters do not vary with time and if heating by the accretion luminosity is neglected.
For a given unstable mode, transition to runaway occurs when fluid in the gain region 
reaches positive specific energy. 
We find approximate instability criteria that accurately 
describe the behavior of the system over a wide region of parameter space. 
The threshold neutrino luminosities are in general different than the limiting value for a steady-state 
solution. 
We hypothesize that multidimensional explosions arise from the excitation of unstable large-scale modes 
of the turbulent background flow, at threshold luminosities that are lower than in the laminar case.
\end{abstract}

\maketitle

\section{Introduction}

Following collapse of the core in a massive star, a hydrodynamic
shock is launched when the central region reaches nuclear density. Significant
energy losses due to neutrino emission and dissociation of infalling nuclei
drain this shock of thermal energy and cause it to stall (e.g., \citealt{bethe90}). 
For slowly-rotating progenitors, the re-absorption of a small fraction of neutrinos 
carrying away the gravitational binding energy of the protoneutron star is thought 
to re-energize this shock and trigger an explosion 
\citep{bethe85}. However, ab-initio radiation-hydrodynamic simulations in spherical
symmetry fail to produce 
this outcome for stars that form iron cores \citep{liebendoerfer01,rampp02,thompson03,sumiyoshi05}.
Success is only obtained for progenitors at the lowest end of the mass range leading to core-collapse 
($\sim8-10M_\sun$, \citealt{kitaura06,burrows07c}).

Increasing the dimensionality of simulations improves conditions for explosion by enabling non-spherical 
hydrodynamic instabilities.
In axisymmetry (2D), large scale shock oscillations combined with neutrino-driven convection lead to
late-time -- albeit somewhat marginal -- explosions in representative stellar progenitors 
(\citealt{marek09} and references therein). 
A recent three-dimensional (3D) hydrodynamic study found that the extra spatial
dimension could make conditions even more favorable, reducing the neutrino luminosity
needed to start an explosion by $\sim 20\%$ relative
to axisymmetry \citep{nordhaus10a}.
The way in which hydrodynamic processes combine to cause this decrease is not well 
understood at present, however. In particular, a change in the
methods and approximations employed can erase this dimensionality effect, suggesting that  
it may manifest only in some region of physical and/or numerical parameter space \citep{hanke11}.
Given the marginality of 2D models, it is important to identify the conditions under which this 
additional reduction in neutrino luminosity occurs.

The purpose of this paper and its companion is to systematically examine the hydrodynamic 
processes responsible for the transition to explosion in a stalled core-collapse supernova shock.
As a first step, we study here the spherically symmetric case. Taking advantage of the
simplicity of the flow geometry and borrowing tools from stellar pulsation theory, we develop
a framework for identifying the relevant processes involved. The spherically symmetric case is also related
(in a time-averaged sense) to three-dimensional models that experience small scale convection 
with no significant large scale shock deformations (e.g., \citealt{nordhaus10a,wongwathanarat2010}).

Spherically symmetric explosions, arising from boosted neutrino luminosities,
involve a transition from a stalled configuration into runaway expansion on timescales longer
than the dynamical time. In a realistic setting, the problem
is fully time-dependent, and a complete analysis must include the mutual feedback of several effects 
(e.g., \citealt{janka01}).
As a way to facilitate the analysis, \citet{burrows93} approximated the problem as a steady-state accretion flow, 
and identified a set of control parameters that determine it. They conjectured that the transition from accretion to explosion
involved a global instability of the flow, with an associated critical stability surface in the space of 
control parameters. State-of-the-art radiation-hydrodynamic simulations show that multidimensional explosions take
place at times when the background flow changes slowly \citep{burrows07,marek09,suwa09}, making a steady-state 
treatment a reasonable starting point to study dimensionality effects.

Global instabilities of the steady-state flow have been identified in several linear
stability studies \citep{yamasaki05,F07,yamasaki07}.
Using a realistic equation of state (EOS) and weak interactions, \citet{yamasaki07}
found that as the neutrino luminosity is increased, both oscillatory and non-oscillatory modes
become unstable.  
The presence of these modes
can be found in spherically-symmetric simulations dating back to the early days of the delayed neutrino
mechanism (e.g., \citealt{wilson86,burrows95,buras06a,ohnishi06,murphy08,FT09b,nordhaus10a,hanke11}).
The instability mechanism behind these modes, their nonlinear development, and their connection
to traditional diagnostics for explosion conditions are not completely understood at 
present, however.

In particular, \citet{burrows93} found that sequences of steady-state models
that satisfy a constraint on the neutrino optical depth are possible only up to a 
limiting neutrino luminosity, for a given mass accretion rate. This limiting value was found
by \citet{murphy08} to be close to the point where spherically symmetric explosions occur,
and by \citet{yamasaki05} to lie at the critical stability curve in the neutrino luminosity vs. mass
accretion rate plane. These results provide support
for the assumption that explosion is indeed obtained at the limiting steady-state luminosity
(e.g., \citealt{pejcha2011}).
The use of a realistic EOS in the linear analysis yields, however, a flow
that becomes unstable for luminosities  $\sim 30\%$ lower than the limiting value, for a 
specific choice of parameters \citep{yamasaki07}. \citet{FT09b} found, in turn, that unstable 
modes develop into explosions when parametric microphysics and steady-state boundary 
conditions are used in time-dependent simulations. 

A more widely used diagnostic for explosion is the ratio of the advection to heating time in the 
gain region \citep{janka98,thompson00}, which is known to have predictive power in the spherically
symmetric case \citep{thompson05}. If an understanding of the hydrodynamic processes leading to explosion -- in any
number of dimensions -- is to be obtained, then the connection between the limiting steady-state luminosity,
stability in spherical symmetry, and proven explosion diagnostics needs to be clarified.

We attempt to shed light on this issue here by studying the properties of unstable modes of the stalled
supernova accretion shock, their nonlinear
development, and the their transition to a runaway solution.
Our approach involves time-dependent hydrodynamic simulations, using a realistic
EOS and weak interactions. The neutrino radiation field is treated parametrically, as is the 
fluid accreting onto the stalled shock. We focus on the evolution of the flow in between
the neutrinosphere and the shock when the parameters describing this flow
are systematically varied. Such a parametric approach has been employed previously to study 
properties of exploding models and the effects of dimensionality
(e.g., \citealt{janka96,ohnishi06,murphy08}).

This paper is organized as follows. Section 2 describes the physical assumptions and numerical methods
employed. Section 3 revisits the limiting neutrino luminosity of the steady-state solution and
its relation to the neutrino optical depth. Section 4 examines the instability affecting the
post-shock flow in the linear and non-linear phase by 
borrowing analysis tools from stellar pulsation theory. By studying the energetics of the 
flow, the nature of the instability cycle, approximate stability criteria, and saturation mechanisms are identified.
Section 5 probes the validity of the instability criteria over a wide region of parameter space. 
The influence of numerical resolution, boundary conditions, and other parameters is also investigated.
Finally, Section 6 contains a summary of our results and a discussion of the implications for the multidimensional case.
Readers not interested in technical details can start in this section, where references to the relevant figures
and equations is made.

\section{Methods}

\subsection{Physical Model and Approximations}
\label{s:assumptions}

We are interested in hydrodynamic instabilities that mediate the onset
of explosion in a stalled core-collapse supernova shock, when the quantities that
determine the global character of the accretion flow -- mass accretion rate, neutrino luminosity, and protoneutron
star radius -- begin to evolve slowly relative to the instability timescales\footnote{The mass of the protoneutron
star is another fundamental parameter of the flow, but it must vary slowly in neutron-star-forming supernovae.}.
This phase starts around $100-200$~ms after core bounce (e.g., \citealt{liebendoerfer01}).
Prior to that, the system is still undergoing transient evolution. Except for progenitors at the
lighter mass end \citep{kitaura06,buras06b}, simulations with neutrino transport
obtain explosions for representative progenitors only after this transient phase is over \citep{burrows07,marek09,suwa09}.

The background accretion flow then evolves on timescales 
$t_{\rm bkg} \gtrsim 0.1-1$~s (e.g., \citealt{bethe90,liebendoerfer01,buras06a}). 
In contrast, the dynamical time, advection time, and thermal time of the flow in between the 
neutrinosphere and the shock are approximately
\begin{eqnarray}
t_{\rm ff}  & \sim & 2 M_{1.3}^{-1/2}\,r_7^{3/2}\textrm{ ms}\\
t_{\rm adv} & \sim & 10 r_7 v_{r,9}^{-1}\textrm{ ms}\\
t_{\rm th}  & \sim & 20 M_{1.3}\,r_7^{-1}\,Q_{\rm net,21}^{-1}\textrm{ ms},
\end{eqnarray}
where $M_{1.3}$ is the mass enclosed within the shock in units of $1.3M_\sun$, $r_7$ is
the radial distance from the center of mass in units of $10^7$~cm, $v_{r,9}$ the radial
velocity in units of $10^9$~cm~s$^{-1}$, and $Q_{\rm net,21}$ the net specific neutrino energy
source term, in units of $10^{21}$~erg~g$^{-1}$~$s^{-1}$ ($\sim 1$~GeV per baryon). 
These timescales are usually shorter than $t_{\rm bkg}$ by a factor of at least several, 
making a steady-state model a reasonable background state to study global hydrodynamic instabilities 
and the transition to explosion \citep{burrows93}.

In this study we adopt a parametric approach in that we do not compute
the neutrino radiation field and boundary conditions starting from a stellar progenitor.
Instead, we regard the quantities determining those processes as free parameters, and study
the behavior of the system as these parameters are varied. This approach has previously been
used by a number of authors to study the hydrodynamic response of the stalled accretion shock
(e.g., \citealt{BM03,ohnishi06,murphy08,iwakami08,FT09b,nordhaus10a,hanke11}).

In our model, we ignore the interior of the protoneutron star, as we are interested in the dynamics of the
fluid between the neutrinosphere and the shock. Instead, we establish an spherical inner boundary at the
radius of the forming neutron star. A time-independent neutrino flux is imposed at this surface.
We consider only electron-type neutrinos and antineutrinos, as these are the only species that exchange
energy with matter with a significant cross-section (e.g., \citealt{bethe90}). 
For simplicity, we establish a single neutrinosphere at a time-independent radius $R_\nu$,
and assume that neutrinos stream isotropically from this surface. In a more realistic treatment, the 
difference between the energy averaged electron neutrino- and antineutrinosphere radius is of the order 
of $10\%$ (e.g., \citealt{buras06a}).

Both neutrino species are assumed to have a Fermi-Dirac
spectrum with zero chemical potential. To remain close to results from more detailed radiation-hydrodynamic simulations,
we set the neutrino luminosities of both species to be equal, but allow them to have different 
neutrinospheric temperatures (e.g., \citealt{janka01}). This is achieved by allowing the normalization of the neutrino
luminosity to vary freely relative to the neutrinospheric temperature and radius (Appendix~\ref{s:weak_rates}).

The contribution to neutrino heating from the accretion luminosity is not included explicitly.
To first approximation, the amount of heating required to start an explosion depends
only on the total flux incident on the gain region, which lies above the cooling region, so
both contributions can be absorbed by the core luminosity in steady-state. However, instability of the
post-shock flow generates changes in the mass accretion rate, hence a non-trivial feedback
from the accretion luminosity can alter the stability properties. We discuss the implications
of this approximation in \S\ref{s:saturation}.

The mass accretion rate is taken to be constant in time.
Only the point-mass gravity of the protoneutron star is considered, as the self-gravity of the envelope is 
a $\lesssim 10\%$ correction to the dynamics. This gravitating mass remains constant in time.

Starting from a steady-state accretion flow defined as above, we evolve the system for 
different values of the control parameters that set the background solution. In particular,
we study the evolution of sequences of models with different neutrino luminosities,
for fixed values of the mass accretion rate and neutrinospheric radius.
Our goal is to identify the hydrodynamic processes by which a quasi-steady state configuration
transitions into an exploding solution, and to determine the conditions (quantified
by the control parameters) for which this transition occurs.

\subsection{Equations and Coordinate System}

Spherical symmetry is assumed throughout this paper, with the origin at the center of 
the protoneutron star. The time-dependent system is described by the equations of mass, momentum,
energy, and lepton number conservation in spherical polar coordinates, with source terms due to the gravity of a 
point mass $M$ and charged-current weak interactions:
\begin{eqnarray}
\label{eq:mass_conservation}
&&\frac{\partial \rho}{\partial t} + \frac{1}{r^2}\frac{\partial}{\partial r} (r^2\rho v_r) = 0\\
\label{eq:mom_conservation}
&&\frac{\partial v_r}{\partial t} + v_r \frac{\partial v_r}{\partial r} + 
\frac{1}{\rho}\frac{\partial p}{\partial r} + \frac{GM}{r^2} = 0\\
\label{eq:energy_conservation}
&&\frac{\partial (\rho e)}{\partial t} + \frac{1}{r^2} \frac{\partial}{\partial r}(v_r [\rho e + p]) + 
\rho v_r \frac{GM}{r^2} = \mathscr{L}_{\rm net}\\
\label{eq:lepton_conservation}
&&\frac{\partial Y_e}{\partial t} + v_r \frac{\partial Y_e}{\partial r}= \Gamma_{\rm net}.
\end{eqnarray}
We denote respectively by $\rho$, $v_r$, $p$, $G$, $e$, and $Y_e$,
the mass density, radial velocity, total pressure, gravitational constant,
specific energy, and electron fraction. Source terms due to weak interactions
are calculated in tabular form, with details provided in Appendix~\ref{s:weak_rates}. The relevant
contributions are the net rate of heating minus cooling per unit volume $\mathscr{L}_{\rm net}$,
and the net rate per baryon of electron minus positron generation $\Gamma_{\rm net}$. 

The system of equations (\ref{eq:mass_conservation})-(\ref{eq:lepton_conservation}) is closed 
with an EOS that yields the relation $p(\rho,e_{\rm int},Y_e)$,
with $e_{\rm int} = e - v_r^2/2$ the specific internal energy. 
We use the model of \citet{shen1998}, as implemented
by \citet{oconnor2010}\footnote{Available at \url{http://stellarcollapse.org}.}. The high-density
part of EOS does not make a significant difference at the densities considered here, $\rho \lesssim 10^{11}$~g~cm$^{-3}$.
The important components are nucleons, alpha particles, and heavy nuclei in nuclear statistical equilibrium,
supplemented by photons and electron-positron pairs with an arbitrary degree of degeneracy and relativity.

\subsection{Initial Conditions}
\label{s:initial_conditions}

The initial condition for time-dependent simulations consists of a steady-state spherical accretion 
flow with a shock at some radial distance $R_s$ from the center,
obtained by solving 
equations~(\ref{eq:mass_conservation})-(\ref{eq:lepton_conservation})
without time derivatives.
Solutions upstream and downstream of the shock
are connected through the Rankine-Hugoniot jump conditions (e.g., \citealt{landau}).

Conditions upstream of the shock are determined by requiring that the velocity, entropy, and electron
fraction of the supersonic flow have specified values at a fixed radial coordinate $r_{\rm vf}$.
In order to obtain values that are in close agreement with more realistic one-dimensional simulations 
(e.g.,  \citealt{liebendoerfer01,buras06a}),
we set $r_{\rm vf} = 100$~km for all simulations. At this radius, we set the radial
velocity to the local free-fall velocity, the entropy per baryon to $5 k_B$, and the
electron fraction to $1/2$.
These parameters together with the mass accretion rate $\dot{M}$, the gravitating mass $M$, and the shock
radius $R_s$ yield the mass density, internal energy, electron fraction, and velocity upstream of the shock, 
by integrating the steady-state equations from $r_{\rm vf}$ to $R_s$. The sensitivity of our results to
these choices is examined in \S\ref{s:other}.
\begin{figure}
\includegraphics*[width=\columnwidth]{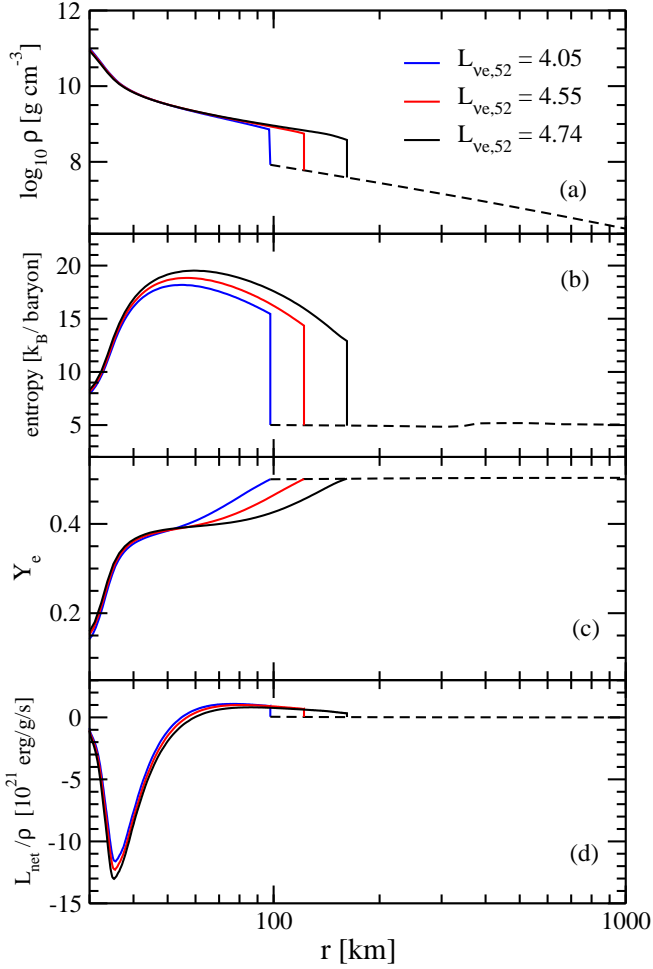}
\caption{Initial profiles of density (a), entropy (b), electron fraction (c), and net neutrino energy generation rate 
$\mathscr{L}_{\rm net}/\rho$ (d) as a function of radius, obtained by solving the steady-state version of
equations~(\ref{eq:mass_conservation})-(\ref{eq:lepton_conservation}) as described
in \S\ref{s:initial_conditions}. Parameters correspond to our fiducial sequence
(\S\ref{s:models}). Curves shown correspond to neutrino luminosities at the approximate instability thresholds
for oscillatory modes (blue) and non-oscillatory modes (red), plus the limiting luminosity for a steady-state
solution (black).}
\label{f:initial_profiles}
\end{figure}

For a given set of parameters, the shock radius $R_s$ is obtained by iteratively solving for the downstream
flow from a trial shock position to $R_\nu$, until an additional constraint is 
satisfied. Following previous studies of steady-state core-collapse supernova flows 
(e.g., \citealt{burrows93,yamasaki05,yamasaki06,yamasaki07,pejcha2011}), this additional closure relation
is obtained by requiring that the neutrino optical depth from $R_\nu$ to the shock 
\begin{equation}
\label{eq:optical_depth_def}
\tau_\nu = \int_{R_\nu}^{R_s} \kappa_\nu \totd r
\end{equation}
is equal to $2/3$.
By default, we set $\kappa_\nu$ to the effective absorption coefficient of electron-type neutrinos
$\kappa_{\rm eff} = \sqrt{\kappa_{\rm abs}(\kappa_{\rm sc}+\kappa_{\rm abs})}$ \citep{janka01}, where 
\begin{equation}
\label{eq:kappa_abs}
\kappa_{\rm abs} \simeq 1.96\times 10^{-7}\, T_{\nu_e,4}^2\, \rho_{10}\,Y_n\textrm{ cm}^{-1}
\end{equation}
corresponds to charged-current absorption and 
\begin{equation}
\label{eq:kappa_sc}
\kappa_{\rm sc} \simeq 0.51\times 10^{-7}\, T_{\nu_e,4}^2\, \rho_{10}\,(Y_n+Y_p)\textrm{ cm}^{-1} 
\end{equation} 
to elastic scattering with nucleons, to lowest order in the neutron-proton mass difference
over the neutrino energy
(e.g., \citealt{bruenn85}). In equations~(\ref{eq:kappa_abs})-(\ref{eq:kappa_sc}), $T_{\nu_e,4}$ is 
the electron neutrinospheric temperature
in units of $4$~MeV, $\rho_{10}$ the mass density in units of $10^{10}$~g~cm$^{-3}$, and $\{Y_n,Y_p\}$ the
number of neutrons and protons per baryon, respectively. To test the sensitivity of our results to this
prescription for the optical depth, we also compute solutions with pure absorption $(\kappa_\nu = \kappa_{\rm abs})$,
or total absorption plus scattering $(\kappa_\nu=\kappa_{\rm abs}+\kappa_{\rm sc})$, thereby bracketing our default 
choice (\S\ref{s:other}).

Figure~\ref{f:initial_profiles} shows characteristic radial profiles of density, entropy, electron fraction, and
net specific neutrino energy generation rate for a few neutrino luminosities, given $\dot{M} = 0.3M_\sun$~s$^{-1}$
and $R_\nu = 30$~km.

\subsection{Time-dependent Implementation}
\label{s:time_dependent}

We use FLASH3.2 \citep{dubey2009} to evolve the system of equations~(\ref{eq:mass_conservation})-(\ref{eq:lepton_conservation}).
The public version of the code has been modified to include the EOS implementation of \citet{oconnor2010},
weak interaction rates, and a grid of variable spacing. 
Details about the latter are provided in Appendix~\ref{s:grid_test}.

We locate the inner simulation boundary at the neutrinospheric radius $R_\nu$.
The outer boundary $r_{\rm max}$ is chosen to be $1000$~km for most models,
corresponding to approximately four times the radius where the nuclear binding
energy of alpha particles equals their gravitational binding energy,
\begin{equation}
\label{eq:r_alpha}
r_\alpha = 254\left( \frac{M_{\rm enc}}{1.3M_\sun}\right)~\textrm{ km},
\end{equation}
where $M_{\rm enc}$ is the mass enclosed at a radius $r_\alpha$.
Above this radius, the shock accelerates significantly during an 
explosion (e.g., \citealt{FT09b} and \S\ref{s:nonlinear}).

We use two types of boundary condition at $R_\nu$. Our default implementation 
fixes all variables in the ghost cells to their initial values, obtained
by continuing the steady-state solution inside $R_\nu$. This prescription
fixes the mass-, momentum-, and energy fluxes leaving the domain. 
A truly reflecting boundary condition would cause the shock to drift outwards
with time due to the accumulation of mass around $R_\nu$, because the latter
does not recede (as it would do with a more realistic treatment, e.g., \citealt{scheck06}). 
\citet{FT09a} and \citet{FT09b} were able to use a reflecting boundary condition
with a fixed neutrinospheric radius
because their parametric cooling function was very centrally concentrated, 
resulting in accumulation of mass in a few cells outside $R_\nu$, and
thus eliminating the shock drift effect. The cooling function used here
has, in contrast, a shallower radial dependence.

To test the influence of the default inner boundary condition on our results, 
we also run a few models that allow an arbitrary amount of mass to leave the domain, 
while still providing pressure support to the accretion flow. 
The density and radial velocity in the ghost cells inside $R_\nu$ are set to
\begin{eqnarray}
\label{eq:outbnd2_dens}
\rho(r)   & = & \rho_0(r) + \rho(r_1) - \rho_0(r_1)\\
\label{eq:outbnd2_velr}
v_r(r) & = & \left(\frac{r_1}{r}\right)^2\,v_r(r_1),
\end{eqnarray}
where $r$ is the radial position of a given ghost cell, $r_1$ corresponds to the center of the first
active cell outside the inner boundary, and the subscript zero labels the
initial steady-state solution. All other variables are set to have zero gradient and are 
thus copied from the innermost active cell into the ghost cells.
The additional factor multiplying the velocity accounts for the radial convergence of 
the flow \citep{ohnishi06}.

The radial cell spacing $\Delta r$ is ratioed (as in, e.g., \citealt{stone92}): $\Delta r_{i+1}/\Delta r_i = \zeta$, 
with $i$ denoting cell number and $\zeta$ some positive real number greater than one. 
We space our cells logarithmically, with $\zeta = (r_{\rm max}/R_\nu)^{N_r}$, where $N_r$ is the total number
of cells, and the size of the cell closest to the inner boundary $\Delta r_{\rm min} = R_\nu (\zeta-1)$ (Appendix~\ref{s:grid_test}).
Guided by convergence tests, we choose either $880$ or $1760$ cells for our runs, although some
models are evolved at resolutions up to $N_r = 3200$ (\S\ref{s:other}). The corresponding values
of $(\zeta-1)$ are $0.4\%$, $0.2\%$, and $0.1\%$, respectively, for $r_{\rm max}=1000$~km and $R_\nu=30$~km.

To avoid problems with the Riemann solver
whenever the internal energy becomes negative,
we implement the prescription of \citet{buras06a}, adding
the nuclear binding energy of alpha particles, heavy nuclei, and
the rest mass energy of electrons to the internal energy during the hydrodynamic
step. The mass fractions of alpha particles and heavy nuclei
are then advected as passive scalars in between calls to
the equation of state. In some cases where the Mach number of the
upstream flow is $\gtrsim 5$, we add an additional zero point of
$7\times 10^{17}$~erg~g$^{-1}$ to
prevent the internal energy to be negative at large radii. This constant
shift makes a negligible difference in the evolution of the shock.

To prevent the system from reaching the lower limit on $Y_e$ allowed by the equation of state 
implementation,  and to account for the decrease in neutrino flux around the 
neutrinosphere in a very crude manner, we multiply all the neutrino source terms by a suppression factor
\begin{equation}
\label{eq:suppression}
f_{\rm sup}(\rho) = e^{-(\rho/\rho_0)} 
\end{equation}
where $\rho_0 = 10^{11}$~g~cm$^{-3}$ is a fiducial density, chosen so as to match the characteristic
density at the neutrinosphere. Given that the effective neutrino optical depth depends chiefly on
density (eq.~[\ref{eq:optical_depth_def}]), the factor in equation~(\ref{eq:suppression})
amounts to a reduction $\sim e^{-\tau_\nu}$ (e.g., \citealt{murphy08}). The choice of $\rho_0$ has some 
mild influence on the threshold for instability, but it doesn't significantly affect the relative differences between
models (\S\ref{s:other}). 

A small initial transient is produced in the form of an outgoing sound wave from the inner boundary and
downgoing entropy and sound waves from the shock. The former acts as an initial perturbation.

\subsection{Models Evolved}
\label{s:models}

We evolve sequences of models with varying neutrino luminosity $L_{\nu_e}$ ($=L_{\bar{\nu}_e}$), for
a given mass accretion rate $\dot{M}$ and prescription for the neutrinospheric radius $R_\nu$. 
Our \emph{fiducial sequence} corresponds to $\dot{M}=0.3M_\sun$~s$^{-1}$ and $R_\nu=30$~km, 
with other parameters fixed to values as described in \S\ref{s:initial_conditions}. This default sequence
closely resembles the flow obtained at late times in the evolution of a $15M_\sun$ progenitor (e.g., \citealt{liebendoerfer01, marek09}).

We explore a wider region of parameter space with a grid of constant $R_\nu$ and $\dot{M}$ sequences
such that $R_\nu=\left\{20,30,40\right\}$~km and $\dot{M}=\left\{0.1,0.3,0.5,0.75,1\right\}M_\sun$~s$^{-1}$,
with our fiducial sequence as one of its members. We add another sequence, following \citet{burrows93}, that
relates the neutrinospheric radius to the neutrino luminosity through the Fermi-Dirac distribution
at constant neutrinospheric temperature and zero chemical potential (equation~[\ref{eq:fnu_normalization}] with 
$N_{\nu_e}=1$), namely $R_\nu \propto L_{\nu_e}^{1/2}$. In this case we use the same grid in  $\dot{M}$ as in the constant 
$R_\nu$ models. 
All sequences are summarized in Table~\ref{t:models}.
\begin{deluxetable}{cccc}
\tablecaption{Model Sequences Evolved\label{t:models}}
\tablewidth{0pt}
\tablehead{
\colhead{$R_\nu$}&
\colhead{$\dot{M}$ ($M_\sun$~s$^{-1}$)} &
\colhead{$L_{\nu_e}$ ($10^{52}$~erg~s$^{-1}$)} &
\colhead{$N_r$} 
}
\startdata
 20 km 			   & 0.1-0.5 & 2.0 - 14.0 &  1760  \\
 30 km 			   & 0.1-1   & 1.5 - 13.7 &  880  \\
 40 km 		           & 0.1-1   & 1.0 - 8.9  &  880  \\
 $\propto L_{\nu_e}^{1/2}$ & 0.1-1   & 1.0 - 5.5  &  880  \\
 30 km 			   & 0.3     & 3.8 - 4.6  &  240-3200 
\enddata
\end{deluxetable}

\section{On the Limiting Neutrino Luminosity of the Steady-State Solution}
\label{s:limiting_luminosity}

\citet{burrows93} found that steady-state solutions to the accretion
shock problem in the supernova context can exist only up to a 
maximum value of the neutrino luminosity (the \emph{Burrows-Goshy limit} hereafter). 
Recently, \citet{pejcha2011} have related this limit to a critical value of the ratio of
sound speed to free-fall speed in the postshock region, in analogy with isothermal accretion flows. 
Here we provide an alternative (but equivalent) explanation.

The existence of a limiting neutrino luminosity for the steady-state solution
is related to the requirement that the neutrino optical depth have a fixed value (eq.~[\ref{eq:optical_depth_def}]).
This closure relation is used for self-consistency with the assumed neutrino radiation field,
but has no significance from the purely hydrodynamic point of view\footnote{The settling of a shocked accretion
flow onto a spherical surface with an arbitrary heating and cooling function is a well-defined hydrodynamic problem,
albeit not necessarily realistic.}.
By relaxing this constraint, one can examine the behavior of $\tau_{\nu}$ as a function of shock
radius and neutrino luminosity, in order to gain insight into the processes that determine this limit.

\begin{figure}
\includegraphics*[width=\columnwidth]{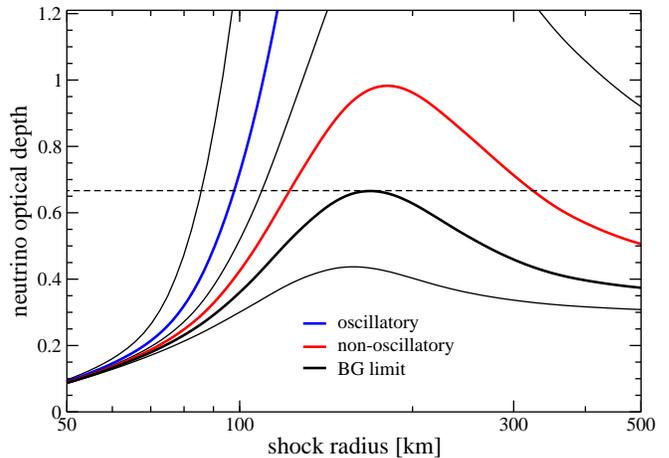}
\caption{Neutrino optical depth $\tau_{\nu}$ (eq.~[\ref{eq:optical_depth_def}]) as a function of shock radius $R_s$,
for parameters corresponding to our fiducial sequence of models (\S\ref{s:models}). Each curve is obtained by fixing
the neutrino luminosity and integrating the time-independent version of equations 
(\ref{eq:mass_conservation})-(\ref{eq:lepton_conservation}) from an arbitrary shock position 
to $R_\nu$, keeping the upstream flow fixed. Curves with decreasing
amplitude correspond to electron neutrino luminosities $L_{\nu_e} = \{3.5,4.05,4.35,4.55,4.74,5\}\times 10^{52}$
erg~s$^{-1}$, respectively. The thick blue and red curves correspond to the approximate instability threshold for oscillatory
and non-oscillatory modes, respectively (\S\ref{s:approximate}). Solutions consistent with neutrino radiation transport 
correspond to the
intersection of each curve with the $\tau_\nu = 2/3$ dashed line. The Burrows-Goshy limit
is marked by a thick black 
curve that has a maximum at exactly $2/3$. The effective optical depth for electron-type neutrinos is used
to compute $\tau_\nu$ (\S\ref{s:initial_conditions}).}
\label{f:optical_depth_curve}
\end{figure}

Figure~\ref{f:optical_depth_curve} shows the neutrino optical depth $\tau_\nu$ as a function of shock radius $R_s$ at $t=0$
for several neutrino luminosities in our fiducial sequence.
Each point along a given curve is obtained by integrating the steady-state version of equations 
(\ref{eq:mass_conservation})-(\ref{eq:lepton_conservation}) from an arbitrary shock position down to a constant 
$R_\nu$, keeping the upstream flow fixed, and calculating equation~(\ref{eq:optical_depth_def}).
For small shock radii, the optical depth is a monotonically increasing function of $R_s$.
For large enough shock radii, however, $\tau_\nu(R_s)$ reaches
a maximum and then decreases for increasing $R_s$. Larger neutrino luminosities move this maximum towards
smaller shock radii and decrease its magnitude. The two points where this curve reaches a value of $2/3$ correspond
to the two physical solutions found by \citet{yamasaki05} and \citet{pejcha2011}. Further increases in the
neutrino luminosity cause the maximum of $\tau_{\nu}$ to fall below $2/3$, making it impossible to satisfy
the optical depth constraint.

\begin{figure}
\includegraphics*[width=\columnwidth]{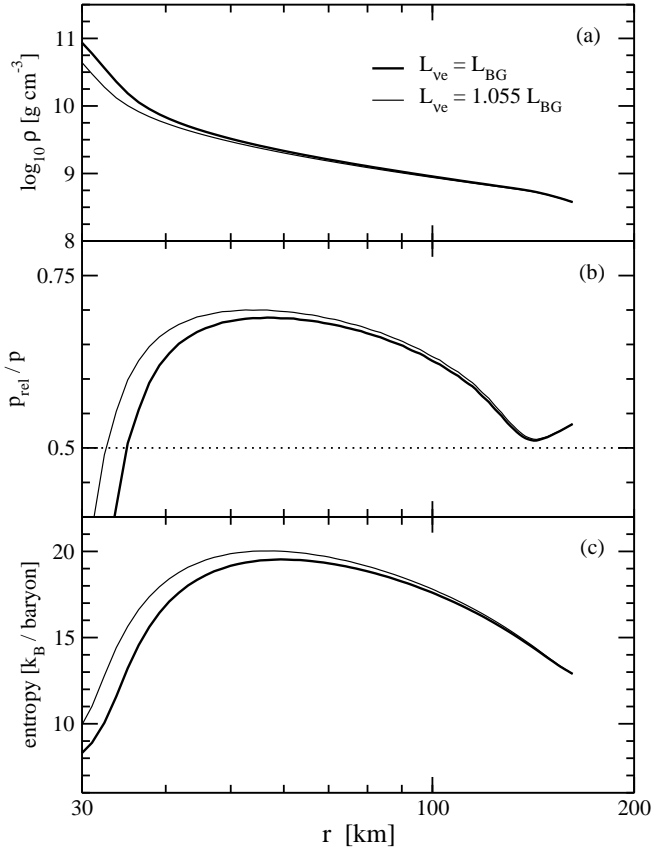}
\caption{Profiles of density (a), ratio of pressure in leptons and radiation to total pressure (b), 
and entropy (c) for two models from our fiducial sequence which have the same shock radius but different 
neutrino luminosities. The thick line corresponds to the Burrows-Goshy limit, while the thin line
has a neutrino luminosity 5\% higher. A higher entropy in the gain region increases the fraction of
the total pressure provided by relativistic particles at fixed radius, softening the density profile and
thus decreasing the density around the neutrinosphere. This is the reason why the lowest curve in
Figure~\ref{f:optical_depth_curve} falls below $\tau_\nu=2/3$.}
\label{f:profiles_taupeak}
\end{figure}

The function $\partial \tau_{\nu}/\partial R_s$ has two contributions. The first comes from the
increase in the volume of the postshock region and is always positive for fixed $R_\nu$. The second
arises from a change in the quantities that determine $\kappa_\nu$, chiefly the density profile 
(eqns.~[\ref{eq:kappa_abs}]-[\ref{eq:kappa_sc}]), 
and can have either sign. 
Figure~\ref{f:profiles_taupeak}a shows the density profile of the model at the Burrows-Goshy limit
and another with neutrino luminosity 5\% larger and the same shock radius (corresponding to the lowest
solid curve in Figure~\ref{f:optical_depth_curve}). The density of the higher luminosity configuration
is lower by a factor of the order of $2$ near the neutrinosphere, where the bulk of the optical depth
is generated.

The density decrease in the cooling layer is caused by an inward motion of the radius at which the
pressure transitions from being dominated by leptons and radiation
to being set by non-relativistic nucleons. 
This is shown by Figure~\ref{f:profiles_taupeak}b. The adiabatic index in the
relativistic particle dominated region is closer to $4/3$, while nucleon domination moves it 
towards $5/3$, causing a steepening of the density profile at this transition\footnote{The steepening
of the density profile with a higher adiabatic index occurs because the temperature varies slowly
with radius near the neutrinosphere. The pressure gradient needed to balance gravity thus arises mostly
from the density gradient. In contrast, in the $\gamma\simeq 4/3$ region the pressure gradient arises mostly
from the changes in the temperature, with the density profile being thus shallower \citep{janka01}.}. 
Inward motion of the transition region relative to $R_\nu$ causes the density to be lower at this radius.
The change in the transition radius in turn is caused by a higher entropy in the gain region 
(Figure~\ref{f:profiles_taupeak}c), which increases the contribution from relativistic
particles.
Finally, the higher entropy is the result of the higher heating rate arising from the higher neutrino 
luminosity.

In the case where the neutrinospheric radius is tied to the neutrino luminosity at
constant temperature (e.g., \citealt{burrows93}, \citealt{yamasaki07}),
an additional contribution arises from the change of the lower limit of integration in equation~(\ref{eq:optical_depth_def}).
This decreases the size of the integration volume for increasing neutrino luminosity relative to the case where $R_\nu$ is
constant, making $\tau_\nu$ smaller than it would otherwise be. 

\citet{pejcha2011} found that the Burrows-Goshy limit can be characterized by a critical value
of the ratio of the sound speed to free-fall speed at some point in the flow, above which it is not
possible to connect a subsonic settling solution to a supersonic accretion flow via the hydrodynamic shock
jump conditions (see also \citealt{yamasaki05}). Implicit in this result, however, is the simultaneous fulfillment of 
the boundary conditions at the upstream side of the shock and at the neutrinosphere when constructing the two solutions. 
Here we have adopted a uni-directional 
approach, keeping only the upstream boundary condition and integrating the fluid
equations inward until an additional constraint is met at the base. 
This allows us to violate the optical depth condition by varying the shock position, which can be
placed at any point in the upstream solution. The conclusions of both approaches regarding the
nature of the  Burrows-Goshy limit are equivalent when viewed in this light, as a higher sound speed 
relative to the free-fall velocity implies higher entropy (see \citealt{murphy11} for a discussion of
the connection between these two quantities through the entropy equation).


\section{Transition to Runaway Expansion}
\label{s:transition_section}

In this section we analyze the instabilities that mediate the transition to
runaway expansion, using simulations results and adapting analysis techniques from
stellar pulsation theory. The term \emph{explosion} implies the existence 
of a star whose envelope is to be ejected by a successful shock. Given the parametric character of our study, we employ 
instead the term \emph{runaway expansion} to denote the \emph{onset} of explosion in a medium with constant
mass accretion rate, as the asymptotic explosion energy is determined by additional processes not included here
(e.g., \citealt{marek09}).

\begin{figure}
\includegraphics*[width=\columnwidth]{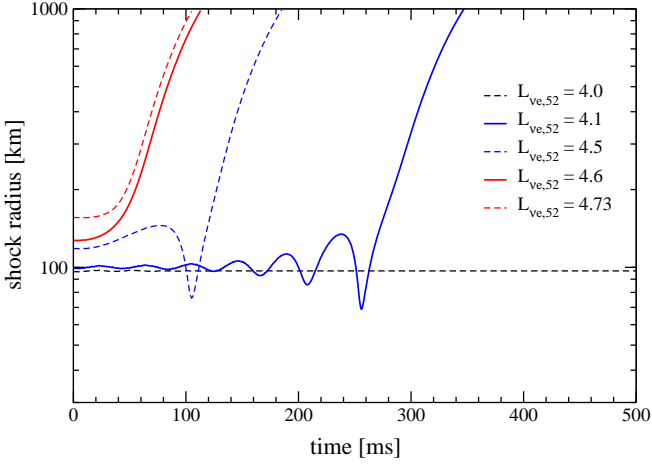}
\caption{Shock radius as a function of time for selected models from our fiducial sequence. Different curves correspond
to different electron neutrino luminosities as labeled ($L_{\nu_e,52} = L_{\nu_e}/10^{52}$~erg~s$^{-1}$). 
Stable, oscillatory, and non-oscillatory modes are plotted as black, blue, and red curves, respectively. 
The maximum luminosity for a steady-state solution at fixed optical depth 
(the Burrows-Goshy limit, \S\ref{s:limiting_luminosity}) corresponds to $L_{\nu_e,52}\simeq 4.74$ 
for this set of parameters.}
\label{f:shock_radius_1d}
\end{figure}

Evolving any of the model sequences described in \S\ref{s:models} yields
a characteristic outcome for the evolution
of the shock radius as a function of time. The result for selected models from
our fiducial sequence is shown in Figure~\ref{f:shock_radius_1d}.
Models with low neutrino luminosity are stable to small perturbations.
For $4.05 < L_{\nu_e}/(10^{52}\textrm{ erg s}^{-1}) < 4.55$, the shock undergoes
oscillations of growing amplitude that transition to runaway
expansion after some time delay. For $L_{\nu_e}> 4.55\times 10^{52}\textrm{erg s}^{-1}$,
initial exponential growth in amplitude occurs without any oscillation, transitioning
into a nearly constant-velocity expansion phase at later times.

The characteristic timescale associated with both types of modes is the
advection time from the shock to the protoneutron star
\begin{equation}
\label{eq:tadv_mass}
t_{\rm adv} = \int_{R_\nu}^{R_s} \frac{dr}{|v_r|} = \frac{M_{\rm env}}{\dot{M}},
\end{equation}
where the second equality is valid when $\dot{M}$ is constant, and 
$M_{\rm env}$ is the mass enclosed between the neutrinospheric radius and the shock.
The fluid velocity is very subsonic, hence the kinetic energy content is small.
Changes in the accretion flow involve changes in the heat content of the fluid through
modulation of the neutrino emission and absorption processes, hence the
instabilities involved are non-adiabatic.
\newpage

\subsection{Work Integral for Accretion Shocks}
\label{s:work_integral}

Any instability involves the conversion of energy from one type to another as the system
transitions into a more favorable configuration. In stellar pulsation theory, the standard tool used
to quantify this conversion is the \emph{work integral} \citep{eddington26,unno89}, 
which is defined as the change in the total energy $E$ of the system over and oscillation cycle
\begin{equation}
\label{eq:W_definition}
W = \oint \totd t \frac{\totd E}{\totd t}.
\end{equation}
In pulsating stars, the total mass of the system is essentially constant
within an oscillation period, thus equation~(\ref{eq:W_definition}) is normally
written to consider Lagrangian changes (e.g., \citealt{unno89})\footnote{The work integral
for stars is also simplified by the pressure falling to very small values
at the surface, which is not the case for flow confined by an accretion shock.}.

In contrast, for an accretion shock neither the mass nor the volume enclosed within the
star and the shock remain constant in time, hence an Eulerian expression must be used if
one wants to adapt this tool to the problem at hand. Writing the total specific energy as 
\begin{equation}
\label{eq:etot_def}
e_{\rm tot} = e - \frac{GM}{r},
\end{equation}
and rearranging the gravitational contribution to equation~(\ref{eq:energy_conservation}), the 
Eulerian rate of change of the total energy in spherical symmetry becomes
\begin{eqnarray}
\label{eq:dEdt_spherical}
\frac{\partial E}{\partial t} & = & \int\totd^3x \frac{\partial (\rho e_{\rm tot})}{\partial t} 
                                  + 4\pi R_s^2 \dot{R}_s\,(\rho e_{\rm tot})\big|_{R_s}\equiv \dot{E}_{\rm tot}\\
\label{eq:dEdt_terms}
                              & = & \dot{E}_{\rm N} - \dot{E}_{\rm up} + \dot{E}_{\rm dn} + \dot{E}_{\rm s},
\end{eqnarray}
with
\begin{eqnarray}
\label{eq:Edot_N}
\dot{E}_{\rm N}  & = & \int_{R_{\rm in}}^{R_s} 4\pi r^2 \totd r\, \mathscr{L}_{\rm net}  \\
\label{eq:Edot_up}
\dot{E}_{\rm up} & = & 4\pi R_s^2 [v_r(\rho e_{\rm tot}+ p)]\big|_{R_s} \\
\label{eq:Edot_dn}
\dot{E}_{\rm dn} & = & 4\pi R_{\rm in}^2 [v_r(\rho e_{\rm tot}+ p)]\big|_{R_{\rm in}}\\
\label{eq:Edot_s}
\dot{E}_{\rm s}  & = & 4\pi R_s^2 \dot{R}_s\,(\rho e_{\rm tot})\big|_{R_s},
\end{eqnarray}
where $\dot{R}_s$ denotes the rate of change of the shock position, and $R_{\rm in}$ is some inner boundary that remains
fixed. 
Equations~(\ref{eq:dEdt_spherical})-(\ref{eq:Edot_s}) can be straightforwardly extended to allow for a moving inner boundary
and non-trivial integration over angles. An analogous expression was derived by \citet{janka01} to study the evolution
of the mass and energy in the gain region.

The first three terms in equation~(\ref{eq:dEdt_terms}) represent, respectively, the integrated neutrino source
terms plus the energy fluxes entering and leaving the integration domain with the accretion flow. 
In steady-state, these three terms cancel
out exactly. The fourth term is non-zero only when the shock moves, and accounts for the change in energy 
caused by a change in the postshock-volume. Because the total specific energy just below the shock is generally
negative due to dissociation losses, this term provides damping on expansion and driving on contraction.

Because the evolution of the shocked flow is not necessarily periodic, we will make use of equation~(\ref{eq:dEdt_spherical})
rather than (\ref{eq:W_definition})
to track the different energetic components in the system. Also, given that the entire flow between $R_\nu$ and
$R_s$ is in sonic contact, analysis of global instabilities must include the cooling region, thus we will 
take $R_{\rm in}$ as close as possible to $R_\nu$.
Details about the calculation of equations~(\ref{eq:dEdt_spherical})-(\ref{eq:Edot_s}) from our simulations are provided 
in Appendix~\ref{s:work_integral_calculation}.

\begin{figure}
\includegraphics*[width=\columnwidth]{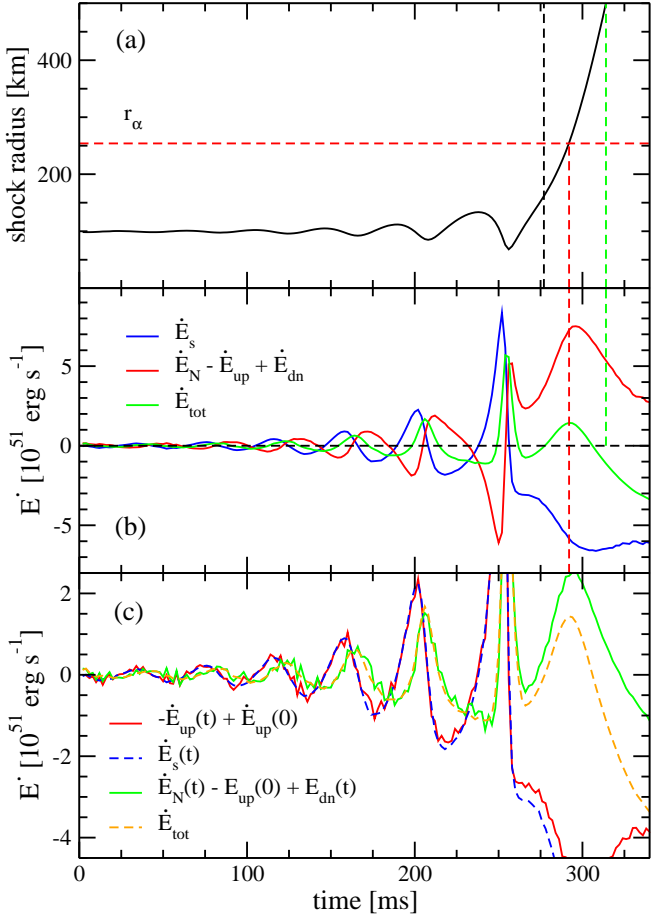}
\caption{Evolution of the different rates of energy change (eq.~[\ref{eq:dEdt_terms}]) for a model in our fiducial 
sequence which explodes via oscillatory instability ($L_{\nu_e,52}=4.1$). 
Panel (a) shows the shock radius as a function of time (solid black), and the radius at which the binding
energy of alpha particles equals their gravitational binding energy (eq.~[\ref{eq:r_alpha}], horizontal red dashed). 
The vertical dashed black line signals the time at which the fluid achieves positive total energy for the first
time (see also Fig.~\ref{f:etot_velx_osc}), while the green dashed line shows the time at which the
sound crossing time from the shock to $R_\nu$ becomes longer than the shock expansion time $R_s/v_s$.
Panel (b) shows the energy generation due to shock motion (blue), the net energy generation from
accretion (red), and the total rate of change of energy (green). 
Panel (c) shows the fluctuation (relative to the initial condition) in the energy flux entering through the shock 
(solid red), the energy generation from shock motion (dashed blue), the fluctuation in the energy flux leaving the 
domain minus neutrino cooling (solid green), and the total rate of change of energy (dashed orange). Note that the 
vertical scale changes relative to panel (b).}
\label{f:dEdt_oscillatory}
\end{figure}

\subsection{Linear Phase: Oscillatory and Non-Oscillatory Modes}
\label{s:linear}

Figure~\ref{f:dEdt_oscillatory} shows the evolution of the shock radius and 
the different terms that account for change of the total energy (eq.~[\ref{eq:dEdt_terms}]), for a 
model that explodes via an oscillatory mode ($L_{\nu_e,52}=4.1$). The lower integration boundary $R_{\rm in}$ 
was chosen to be a few cells above $R_\nu$, to eliminate boundary effects
(Appendix~\ref{s:work_integral_calculation}).

In the linear phase, the magnitude of the energy generation by shock motion $\dot{E}_s$ peaks when the shock velocity
is the largest, and is dominated by the gravitational term in $e_{\rm tot}$. Contractions thus
inject positive energy into the postshock region. The net energy generation from accretion
$\dot{E}_{\rm N}-\dot{E}_{\rm up}+\dot{E}_{\rm dn}$ also becomes non-zero once the system deviates from equilibrium.
Its magnitude is much smaller than the individual terms that comprise it, 
showing that imperfect cancellation between the energy fluxes and neutrino source terms is the additional source of energy.
The lowest panel in Figure~\ref{f:dEdt_oscillatory} shows that in fact the flow adjusts itself in the linear phase to
match the energy generation from shock motion with the excess energy flux entering the domain, 
$\dot{E}_s(t) \simeq \dot{E}_{\rm up}(t)-\dot{E}_{\rm up}(0)$.
The total energy generation can then be accounted for entirely by the deviation from steady-state of
the neutrino source terms and the energy flux leaving the domain toward the protoneutron star.
In other words, oscillatory instability arises from a mismatch in the dissipation of the accretion energy in 
a system with a moving boundary.

\begin{figure}
\includegraphics*[width=\columnwidth]{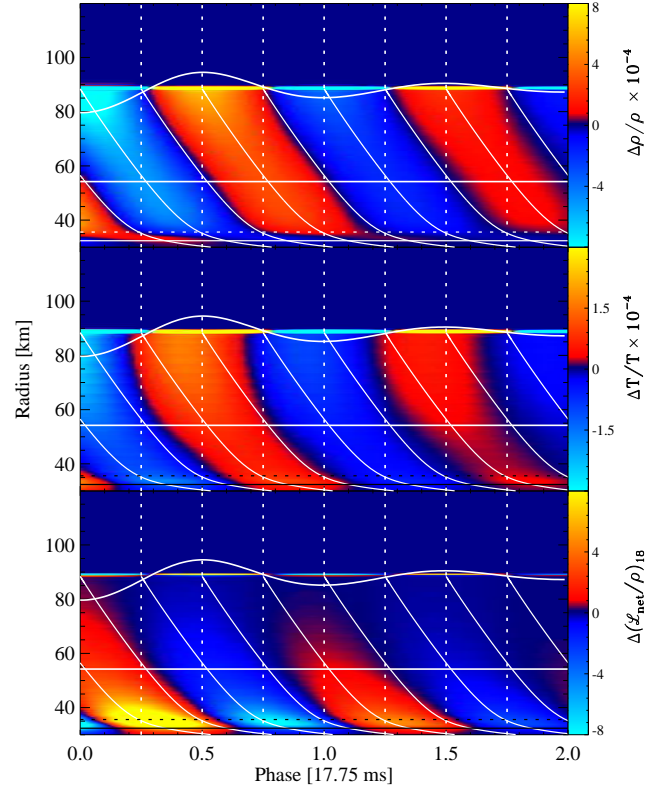}
\caption{Spacetime diagram showing different perturbed quantities as a function of time for a stable model
in our fiducial sequence ($L_{\nu_e,52} = 3.5$). Two complete oscillation cycles are shown, and the differences
are taken relative to values at very late times, where the flow reaches perfect steady-state. The top panel
shows normalized density fluctuations, the middle normalized temperature fluctuations, and the bottom
the fluctuation in the specific net neutrino source term $\mathscr{L}_{\rm net}/\rho$ in units of $10^{18}$~erg~g~s$^{-1}$. 
Horizontal lines correspond to the gain radius (upper solid white), maximum cooling (dashed black/white), and radius where cooling
has decreased by an e-folding from its maximum (white/black solid near bottom of each panel). 
Diagonal lines correspond to streak lines in the unperturbed
solution. The amplitude of the shock radius (white wavy curves) has been increased by a factor of 100 for clarity.}
\label{f:spacetime}
\end{figure}

The origin of this imbalance can be traced
back to the steep temperature dependence of the neutrino cooling function, 
$\mathscr{L}_{\rm C}/\rho \propto T^6$ (eq.~[\ref{eq:L_C_def}]).
Figure~\ref{f:spacetime} shows that the perturbation to the specific neutrino source term closely mirrors
the temperature perturbation, with the opposite sign (after removing the density dependence, the heating
term depends only on the neutron and proton abundance, linearly). Thus an increase in the temperature from
its steady-state value causes an increase in the rate of cooling and hence excess energy dissipation. 
This results in contraction and cooling of the postshock volume on a sound crossing time, reversing the cycle when the 
lower temperature fluid reaches the cooling region by advection.

Figure~\ref{f:spacetime} also shows that the period of the oscillation is related to the advection time from the 
shock to a radius near the point of maximum cooling. At this radius the flow achieves maximum deceleration,
and advected perturbations are efficiently converted into acoustic waves \citep{scheck08,foglizzo09,sato09}.
The normalized density perturbation is significant only outside of this radius, below which it undergoes a phase shift. 
The model shown in Figure~\ref{f:spacetime} has a period very close to twice the advection timescale from the shock to this
radius. This is very close to the period of an oscilltory mode with no heating \citep{FT09a}. As the flow becomes unstable,
however, the oscillation period becomes increasingly longer, achieving nearly four advection times at the threshold for oscillatory
instability. This lengthening of the oscillation period with neutrino luminosity has been attributed to buoyancy effects 
by \citet{yamasaki07}.

\begin{figure}
\includegraphics*[width=\columnwidth]{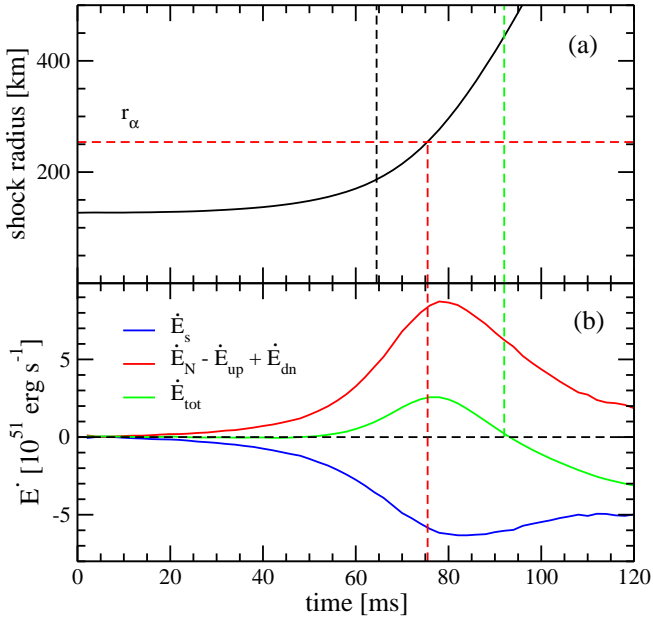}
\caption{Same as Figure~\ref{f:dEdt_oscillatory}, but for a model that explodes via the non-oscillatory
instability ($L_{\nu_e,52}=4.6$). Only the two upper panels are shown, as the fluctuation in $\dot{E}_{\rm up}$
does not compensate $\dot{E}_s$, as in the last expansion of the oscillatory mode.}
\label{f:dEdt_nonoscillatory}
\end{figure}

Non-oscillatory modes grow exponentially in amplitude and transition directly into runaway expansion.
The different terms that make up the rate of change of total energy are shown in Figure~\ref{f:dEdt_nonoscillatory}
(the model has $L_{\nu_e,52} = 4.6$). The evolution of these terms resembles the last expansion phase of oscillatory modes, 
where the shock motion term acts only as a sink of energy and is not matched by the deviation in $\dot{E}_{\rm up}$ from steady-state. 
The expansion is driven by the net energy generation from accretion, which is positive throughout. 
 
\subsection{Approximate Instability Criteria}
\label{s:approximate}

The only way to obtain an exact instability threshold for these non-adiabatic modes is to perform a linear 
perturbation analysis of equations~(\ref{eq:mass_conservation})-(\ref{eq:lepton_conservation})
such as that done by \citet{yamasaki07}. 
In the absence of such a calculation, we have instead searched for physically motivated instability criteria 
that correctly describe the behavior of our simulations over a wide range of parameter space. We discuss
here their definition, and leave for \S\ref{s:relation} their comparison with simulations over an extended
region of parameter space.

For oscillatory modes, we have found that instability sets in when the advection time through the gain region 
is longer than the time required to advect from the gain radius to a point close to the node in the 
density perturbation (Figure~\ref{f:spacetime}) times some coefficient close to unity. This is equivalent
to requiring more mass to reside in the gain region than in the part of the cooling region above the
node in the density perturbation, at any given instant (eq.~[\ref{eq:tadv_mass}]). 
To obtain a coefficient
of unity, the relevant location is the radius at which the density and pressure perturbations are
180 degrees out of phase with each other. In the steady-state solution, and for most of our models, this
position is approximately the radius closest to $R_\nu$ where the specific cooling has decreased in magnitude 
by one e-folding from its maximum (e.g., Figure~\ref{f:initial_profiles}d).
 
\begin{figure}
\includegraphics*[width=\columnwidth]{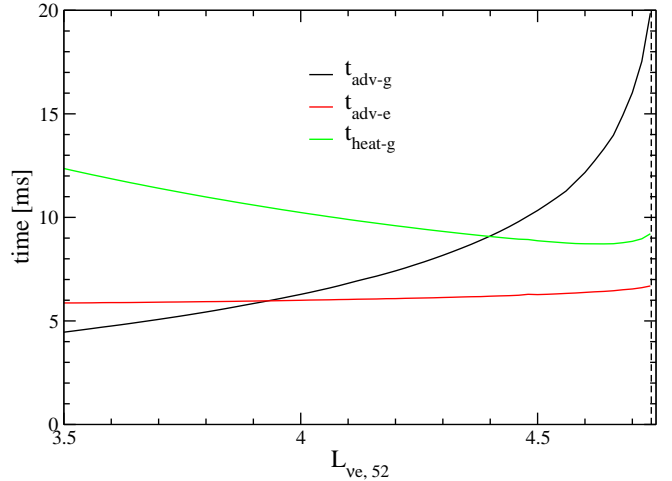}
\caption{Characteristic timescales that determine the approximate instability condition for oscillatory and
non-oscillatory modes, as a function of electron neutrino luminosity in our fiducial sequence 
(compare with Figure~\ref{f:shock_radius_1d}). Shown are the advection time
over the gain region (black, eq.~[\ref{eq:tadv_gain}]), advection time from the gain radius to the point where cooling has decreased
by an e-folding from its maximum (red, eq.~[\ref{eq:tadv_coole}]), and timescale to change the total energy in the gain
region by neutrino heating (green, eq.~[\ref{eq:theat_gain}]). The vertical dashed line marks the Burrows-Goshy limit.}
\label{f:instability_criteria}
\end{figure}

The approximate instability criterion is then 
\begin{equation}
\label{eq:instability_osc}
t_{\rm adv-g} > t_{\rm adv-e},
\end{equation}
with 
\begin{equation}
\label{eq:tadv_gain}
t_{\rm adv-g} = \int_{R_g}^{R_s}\,\frac{\totd r}{|v_r|}
\end{equation}
and
\begin{eqnarray}
\label{eq:tadv_coole}
t_{\rm adv-e} & = & \int_{r_e}^{R_g}\,\frac{\totd r}{|v_r|}\\
\label{eq:rcoole_def}
\left(\frac{\mathscr{L}_{\rm net}}{\rho}\right)\bigg|_{r_e} & = & 
\frac{1}{e}\left(\frac{\mathscr{L}_{\rm net}}{\rho}\right)\bigg|_{\rm min},
\end{eqnarray}
and where the solution to equation~(\ref{eq:rcoole_def}) closest to $R_\nu$ is taken.
The dependence of these two timescales on neutrino luminosity
is shown in Figure~\ref{f:instability_criteria}.

It is worth emphasizing that $r_e$ was chosen because it yields a coefficient of unity. Another location
close to the node in the density perturbation could also have been used, with a slightly different coefficient. For example,
taking the radius of maximum specific cooling $r_{\rm cmax}$ yields an instability condition $t_{\rm adv, gain} > 1.5 t_{\rm cmax}$, 
where the latter timescale is computed by replacing $r_e$ with $r_{\rm cmax}$ in equation~(\ref{eq:tadv_coole}). 

A physical justification for $r_e$ can be found by looking at the behavior of the relative phase between the
pressure and density perturbations. It is a general result from stellar pulsation theory that, for oscillatory instability,
an excitation mechanism is required which causes the system to be absorbing heat at the point of maximum compression 
(e.g., \citealt{cox74}). This way, the pressure maximum is reached during the expansion phase, leading to positive work
over an oscillation cycle, and hence to a net increase in the pulsation kinetic energy. In the system under study, such
a phase lag of the pressure perturbation relative to the density is satisfied below the node of the density perturbation.
The point where both perturbations are 180 degrees apart is approximately $r_e$ for most of our models, with the pressure leading the 
density at any given radius above it\footnote{To determine the precedence of density or pressure maxima, we take two peaks that are
closest in phase. This is only possible when perturbations have a phase shift different than 0 or 180 degrees.}.  
In cases where $r_e$ does not trace the phase radius correctly, the criterion in
equation~(\ref{eq:instability_osc}) loses accuracy. The relation between the phase radius and the cooling profile
is most likely a consequence of the way we suppress the neutrino source terms with density (eq.~[\ref{eq:suppression}]);
different implementations of neutrino transport will most likely find slightly different values.

An interesting question is how the phase lag criterion for stars can be adapted to the problem at hand, given that fluid elements 
are advected downstream instead of returning to their original position. Taken at face value, the phase lag criterion would 
dictate that the driving region resides below the node of the density perturbation. However, the pressure lags the density
in this region for both stable and unstable models, and $r_e$ hardly changes over a wide range in $L_{\nu_e}$ 
(Figure~\ref{f:instability_criteria}). Such question is likely to be at the root of the instability mechanism of long wavelength
modes of the Standing Accretion Shock Instability (SASI, \citealt{BM03}) with heating included, and will not be further
pursued in this paper.

In the case of non-oscillatory modes, the instability threshold is very close to the point where 
the advection time through the gain region is longer than the time required to change the total
energy through heating,
\begin{equation}
\label{eq:instability_nonosc}
t_{\rm adv-g} > t_{\rm heat-g}
\end{equation}
with
\begin{equation}
\label{eq:theat_gain}
t_{\rm heat-g} = \frac{\int_{R_g}^{R_s}\totd^3 x\, (\rho e_{\rm tot})}{\int_{R_g}^{R_s}\totd^3 x\, \mathscr{L}_{\rm net}}.
\end{equation}
The dependence of $t_{\rm heat-g}$ on neutrino luminosity is also shown in Figure~\ref{f:instability_criteria}.
This relation has long been known to provide a predictive criterion for runaway expansion in spherically symmetric 
core-collapse simulations \citep{thompson05,buras06b,murphy08,marek09}. Its interpretation is straightforward: neutrino heating
is effective enough to change the internal energy of the flow during its transit time through the gain region.
As a consequence, the pressure changes significantly and the shock re-adjusts to a new
position \citep{janka98}. 

Note that equation~(\ref{eq:instability_nonosc}) is a global condition on the gain region and includes the gravitational
binding energy, differing from
local definitions of this ratio such as those in \citet{thompson05} and \citet{pejcha2011}. The fluid
does not achieve positive energy over a  time $t_{\rm adv-g}$ ($\sim 10$~ms, Figure~\ref{f:instability_criteria}),
but instead takes a multiple of this timescale to transition into runaway expansion.
As we discuss in the next section, onset of runaway occurs when the fluid achieves positive energy
for the first time.

We have checked whether any instability threshold can be described by a fixed
value of the parameter $\chi$ that measures the effects of buoyancy in an advective flow,
\begin{equation}
\chi = \int_{R_g}^{R_s} |\omega_{\rm bv}| \frac{\totd r}{|v_r|},
\end{equation}
where $\omega_{\rm bv}$ is the Brunt-V\"ais\"al\"a frequency,
\begin{equation}
\omega_{\rm bv}^2 = \frac{GM}{r^2}\left(\frac{1}{\Gamma_s}\frac{\partial \ln p}{\partial r } - \frac{\partial \ln \rho}{\partial r} \right),
\end{equation}
with $\Gamma_s = (\rho/p)\, c_s^2$. When $\chi \gtrsim 3$, buoyancy is expected to overcome the 
stabilizing effect of advection and lead to convective instability in the multidimensional case \citep{foglizzo06}.
Even though all of our sequences have $\chi$ in the range 1-10 near the threshold for instability, in none of them
does this parameter have a constant value along the critical stability curves. For example, in our fiducial sequence,
the value of this parameter at the non-oscillatory threshold ranges from $\chi \simeq 7$ for $\dot{M} = 0.1M_\sun$~s$^{-1}$ 
to $\chi = 1.7$ at $\dot{M} = 1M_\sun$~s$^{-1}$. It is worth pointing out that the analysis of \citet{foglizzo06}
applies to infinitesimal perturbations. The results of \citet{scheck08} show that the SASI can
trigger convection through finite amplitude density fluctuations in cases where $\chi < 3$, so our findings
do not necessarily imply that convection will be suppressed at large accretion rates in the quasi steady-state
approximation (\S2.1).

\subsection{Nonlinear Phase and Runaway Expansion}
\label{s:nonlinear}

Once oscillatory modes become unstable, the amplitude grows steadily until 
the oscillation cycle is broken and the system transitions into runaway expansion. 
Figure~\ref{f:dEdt_oscillatory} shows that not only the shock radius but also the different
energy generation terms undergo a qualitative change in behavior relative to the linear phase
once this point is reached. In particular, the energy generation due to shock motion decouples from the fluctuation
in the energy flux entering though the shock. Expansion is driven by the net energy generation from accretion,
with damping from the shock motion term. 

\begin{figure}
\includegraphics*[width=\columnwidth]{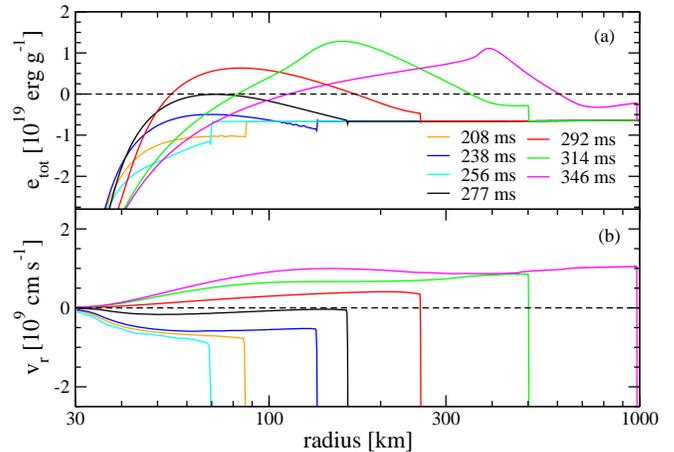}
\caption{Total specific energy $e_{\rm tot}$ (a) and radial velocity (b) as a function of radius
for the model shown in Figure~\ref{f:dEdt_oscillatory}, which explodes via the oscillatory instability.
Different curves correspond to different times in the simulation, as labeled. The black curve (277 ms),
red curve (292 ms), and green curve (314 ms) correspond to the vertical dashed lines of the same color in 
Figure~\ref{f:dEdt_oscillatory}a.}
\label{f:etot_velx_osc}
\end{figure}

A physical origin for this transition point
can be found by inspecting the evolution of the total specific
energy and radial velocity in the system. Figure~\ref{f:etot_velx_osc} shows that this time
corresponds approximately to the point where the specific energy behind the shock becomes positive for the first
time. This time is marked by a vertical black dashed line in Figure~\ref{f:dEdt_oscillatory}a and the snapshot
at $277$~ms in Figure~\ref{f:etot_velx_osc}.
At about the same time, the fluid just inside the shock achieves
positive radial velocity, reversing the accretion flow. This transition point is equivalent to the
\emph{escape temperature} condition discussed in \citet{burrows95}.

Physically, the above result is dependent on the definition of the zero points of energy in the context of a Newtonian
framework.
The EOS used here takes this zero point to be, per baryon, the atomic mass unit \citep{shen1998}. Our expression for the total 
specific energy, equation~(\ref{eq:etot_def}), takes the zero point of gravitational binding energy at an infinite 
distance from the central mass. Given these definitions, the condition of positive energy leading to runaway
expansion is a well-defined mathematical concept. A self-consistent determination of this condition would require
inclusion of the rest mass energy and gravitational field calculated self-consistently in a general relativistic framework. 

The shock accelerates when it approaches the radius where the binding energy of alpha particles
equals their gravitational binding energy (eq.~[\ref{eq:r_alpha}]). This can be understood from
the fact that the dissipation due to shock motion $\dot{E}_s$ decreases in magnitude with increasing
radius, because the total specific energy behind the shock becomes less negative. 
This can be seen in Figure~\ref{f:dEdt_oscillatory}b,
where the blue curve reaches a minimum value. By now most of the gain region has reached positive energy, 
except for a narrow layer behind the shock, and most of the fluid has positive velocity, effectively becoming
a wind-like solution.
\begin{figure}
\includegraphics*[width=\columnwidth]{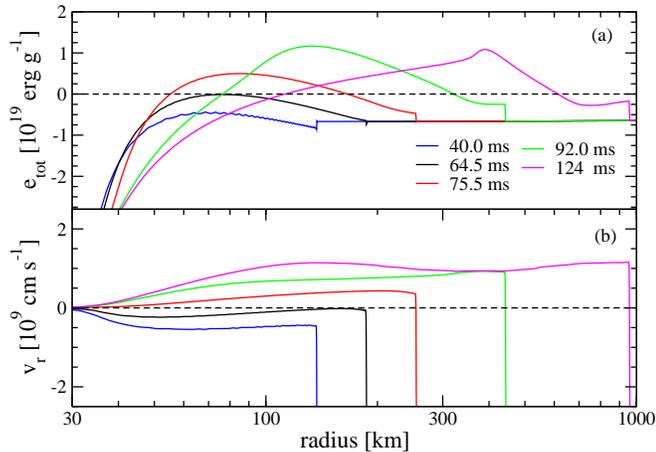}
\caption{Same as Figure~\ref{f:etot_velx_osc}, but for the model shown in Figure~\ref{f:dEdt_nonoscillatory}, which
explodes via the non-oscillatory instability. Black (64.5 ms), red (75 ms), and green (92 ms) curves correspond
to the times signaled by vertical dashed curves of the same color in Figure~\ref{f:dEdt_nonoscillatory}.}
\label{f:etot_velx_nonosc}
\end{figure}

A final transition occurs when the shock begins to mechanically decouple from the
protoneutron star atmosphere. This occurs approximately at a time when the sound crossing
time from the shock to $R_\nu$ becomes longer than the shock expansion time $R_s/v_s$, where $v_s$
is the shock velocity. This time is marked as a vertical green dashed line in Figure~\ref{f:dEdt_oscillatory}a
and corresponds to the curve at 314 ms in Figure~\ref{f:etot_velx_osc}. 
This mechanical decoupling explains the fact that the net energy generation becomes negative yet
the shock continues to expand, as inferred from Figure~\ref{f:dEdt_oscillatory}. It is worth noting
that velocities below the shock are subsonic throughout the time period considered here.

Non-oscillatory modes transition directly into runaway expansion, and their evolution closely
resembles that of oscillatory modes after reaching positive energy. The corresponding snapshots
of total specific energy and radial velocity at different phases are shown in 
Figure~\ref{f:etot_velx_nonosc}.

In all of the sequences simulated, transition to runaway expansion is achieved after the flow
becomes radially unstable. We have found no unstable mode that saturates, nor any model that having
achieved positive energy, failed to continue into runaway expansion. Our preliminary conclusion is that
for the set of assumptions adopted in this paper, radial instability is a sufficient condition for
transitioning into runaway expansion. 

\subsection{Conditions for Saturation}
\label{s:saturation}

There are many examples in the literature where one-dimensional stalled shocks start to 
expand or oscillate, but then fizzle (e.g., Figure 1 of \citealt{janka96}). It is thus imperative to identify
the processes that can lead to saturation, as our preliminary finding linking instability to runaway may break 
down in some region of parameter space.

Insight on this question can be obtained again by inspecting
the different processes that contribute to the change in the total energy
in the post-shock region (eq.~[\ref{eq:dEdt_terms}]).
What we are interested in are dissipation processes that contribute
with negative energy generation.

In the present study, we are ignoring evolution of the neutrinospheric parameters ($R_\nu$,
core luminosities, and spectra) and heating due to the accretion luminosity, hence
the dominant dissipation mechanism arises from the change in the post-shock volume, $\dot{E}_s$, as
inferred from Figures~\ref{f:dEdt_oscillatory} and \ref{f:dEdt_nonoscillatory}.
This term is made up of three factors: (1) the specific energy below the shock, (2) the density profile, 
and (3) the rate of change of the post-shock volume. 
\begin{enumerate}

\item At the typical radii where supernova shocks stall ($100-200$~km), the total specific energy $e_{\rm tot}$
      is usually negative. Because the temperature decreases outward, nucleons below the shock
      recombine first into alpha particles and then heavy nuclei as the shock moves out, yielding an increase in the thermal energy
      that is comparable to the gravitational binding energy \citep{bethe96,FT09b}. This causes the total specific
      energy to become significantly less negative, decreasing the size of $\dot{E}_s$ as the shock expands. This
      can be most clearly seen in Figure~\ref{f:dEdt_oscillatory}b, which shows a clear transition in the evolution
      of $\dot{E}_s$ when the shock reaches $r_\alpha$.
      The results of \citet{FT09b} show that using a constant dissociation energy indeed quenches runaway
      expansion, in direct contrast to allowing alpha particles to recombine, because in the first case the dissociation energy becomes
      an increasingly larger fraction of the local gravitational binding energy as the shock expands. Even then,
      saturation occurs at a radius that is several times the initial shock radius. Except for unrealistically
      small shock stagnation radii ($\lesssim 50$~km), this saturation channel is unlikely to be of importance.

\item The evolution of the density profile below the shock and the mass accretion rate are
      tied to the density profile of the progenitor
      at the onset of collapse. For an iron core supported by relativistic electrons the
      density scales like $r^{-3}$, yielding a mass accretion rate that scales inversely
      with time at fixed radius when mass conservation and a near free-fall velocity field
      are assumed \citep{bethe90}. A time-independent mass accretion rate corresponds to
      a progenitor density profile $\propto r^{-3/2}$, which would be given by an adiabatic index of $5/3$ 
      in hydrostatic equilibrium. In other words, a time-independent mass accretion rate, as
      assumed in our sequences, is already unrealistically high and overestimates 
      the dissipation.

\item Because $\dot{E}_s$ is negative on expansion, it tends to 
      stabilize the shock velocity for a given energy dissipation rate.
      The dominant energy source for expansion is neutrino heating.
      As long as the total heating remains larger than $\dot{E}_s$ before the shock
      is completely decoupled from the atmosphere, the shock will 
      continue to expand.

\end{enumerate}

We therefore conclude that for the physical assumptions of the present study, \emph{radial instability
is a sufficient condition for runaway expansion}. 
\begin{figure*}
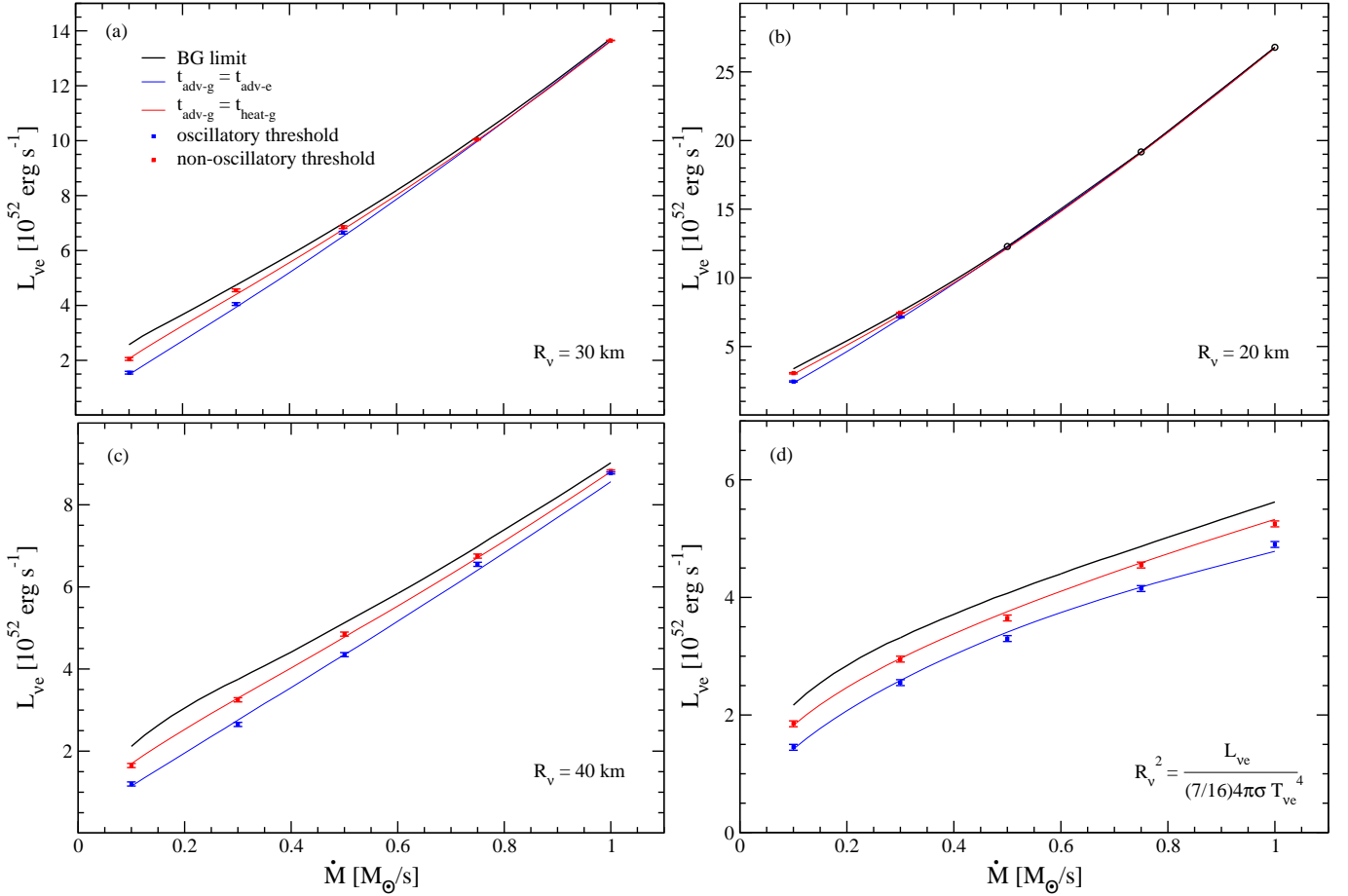

\includegraphics*[width=0.5\textwidth]{f11a.eps}
\includegraphics*[width=0.5\textwidth]{f11b.eps}
\includegraphics*[width=0.5\textwidth]{f11c.eps}
\includegraphics*[width=0.5\textwidth]{f11d.eps}
\caption{Threshold electron neutrino luminosity for radial instability as a function
of mass accretion rate, for different sequences of models. Panels (a), (b), and (c) correspond
to sequences with fixed neutrinospheric radii at $R_\nu=\{30,20,40\}$~km, respectively, while the
sequence in panel (d) relates $R_\nu$ to the neutrino luminosity via the black body relation (\S\ref{s:models}). 
Data points denote results from time-dependent simulations, where the threshold for explosion is taken as
the average between stable and unstable models, with the error bars showing the size of the difference in $L_{\nu_e}$
($\Delta L_{\nu_e} < 10^{51}$~erg~s$^{-1}$). Blue points correspond to oscillatory and red to non-oscillatory 
instability, respectively.
Curves correspond to our approximate instability criteria for oscillatory (blue, eq.~[\ref{eq:instability_osc}]), 
and non-oscillatory modes (red, eq.~[\ref{eq:instability_nonosc}]). Black curves correspond to the limiting luminosity
of the steady-state configuration at constant optical depth, calculated as described in \S\ref{s:limiting_luminosity}.
Black circles in panel (b) denote the absence of instability (and thus explosion) for $\dot{M}\ge 0.5M_\sun$~s$^{-1}$ for 
all luminosities up to and including the Burrows-Goshy limit.}
\label{f:BG93_data}
\end{figure*}

In the general case, however, evolution of the neutrinospheric parameters and neutrino heating
by the accretion luminosity can provide significant energy dissipation and saturate
the instability. By inspection of equation~(\ref{eq:dEdt_spherical}), one can find three sources
of negative energy.

First, evolution of the core neutrino flux affects the energetics via $\dot{E}_{\rm N}$. 
If the net effect of decreasing neutrino luminosities and increasing mean neutrino energies is a 
decrease in the energy deposition in the gain region, the instability can be suppressed in two ways.
The linear instability depends on the extent of the gain region through $t_{\rm adv,gain}$, hence
a rapid decrease of heating can stabilize oscillations or non-oscillatory expansion if the flow
has not yet achieved positive energy when the stability criterion is reversed. Second, a decrease
in the rate of neutrino energy deposition can lead to quenching of the runaway phase if the total 
rate of heating fails to keep up with dissipation due to $\dot{E}_s$. The results of \citet{janka96} are
consistent with the operation of this saturation channel.
 
Second, the contribution of the accretion luminosity to the total neutrino flux in realistic models is
$\sim 50\%$ (e.g., \citealt{liebendoerfer01}). The important aspect to keep in mind here
is that shock expansion relative to its steady-state position leads to a decrease in the magnitude
of the mass accretion rate. Perturbing the mass conservation equation yields \citep{F07}
\begin{equation}
\frac{\delta\dot M}{\dot{M}} = \left(\frac{1}{v_2}-\frac{1}{v_1} \right)\,\delta v_s,
\end{equation}
where $v_1$ and $v_2$ are the (negative) upstream and downstream fluid velocities, and $\delta v_s$ is the
shock velocity perturbation. Taking $\delta v_s$ real and positive (shock expansion) yields $\delta \dot{M} / \dot{M} < 0$,
or a decrease in the magnitude of the accretion rate.
Hence including the heating from accretion neutrinos adds a non-trivial feedback to both oscillatory and non-oscillatory 
modes, altering the instability criteria. \citet{buras06a} find that indeed spherical oscillations are damped by 
the decrease in the heating due to the dropping mass accretion rate.
As noted by a number of previous works, the runaway expansion phase also cuts off accretion in spherical
symmetry. A larger core neutrino flux is therefore required to sustain expansion, relative to the 
light bulb heating case,  and the causality relation between radial instability and explosion we have
found here will be violated.

Finally, the contraction of the protoneutron star generates in itself a sink of energy
in the post-shock region. Including this effect, and keeping everything else constant, would add a term of the form
\begin{eqnarray}
\dot{E}_{\nu{\rm sph}} & = & -4\pi R_\nu^2 \dot{R}_\nu (\rho e_{\rm tot})_{R_\nu}\\
  	     & \simeq & -10^{50} R_{\nu,30}^2\,\dot{R}_{\nu,6}\, \rho_{11}\, e_{{\rm tot},19}\textrm{ erg s}^{-1}
\end{eqnarray}
to equation~(\ref{eq:dEdt_spherical}), with $R_{\rm in}=R_\nu$. The notation in the second
equality is $R_{\nu,30}=R_\nu/(30\textrm{ km})$, $\dot{R}_{\nu,6} = \dot{R}_\nu/(-10^6\textrm{ cm s}^{-1})$,
$\rho_{11} = \rho/(10^{11}\textrm{ g cm}^{-3})$, and $e_{{\rm tot},19} = e_{\rm tot}/(-10^{19}\textrm{ erg g}^{-1})$.
In contrast to $\dot{E}_s$, contraction leads to energy dissipation.
This quantity is smaller than typical core neutrino luminosities, but approaches the net energy generation
terms that regulate instabilities (\S\ref{s:linear}). A more careful analysis would 
need to include the additional cooling and heating by accretion neutrinos generated by enlarging the
post-shock cavity.
The one-dimensional results of \citet{janka96} show that indeed $\sim 10\%$ higher core neutrino luminosities are required 
to start an explosion in models that experience protoneutron star contraction relative to fixed cores, 
when all other parameters are similar.

\section{Instability Thresholds and Relation to the Limiting Steady-State Luminosity}
\label{s:relation}

\subsection{Dependence on Mass Accretion Rate and Neutrinospheric Radius}

For each of the simulation sequences described in \S\ref{s:models}, we have searched for 
the instability thresholds of oscillatory and non-oscillatory modes. Figure~\ref{f:BG93_data} shows the
resulting threshold luminosities as a function of mass accretion rate. In 
all cases, data points correspond to the average luminosity between two models at both
sides of the instability threshold. The separation in luminosity, shown as error bars,
is less than $10^{51}$~erg~s$^{-1}$.

Figure~\ref{f:BG93_data} also shows the limiting luminosity for a steady-state configuration 
(the Burrows-Goshy limit), calculated as described in  \S\ref{s:limiting_luminosity}. 
The resulting curve agrees qualitatively with the results of \citet{burrows93}, \citet{yamasaki05,yamasaki06}, 
and \citet{pejcha2011}.
Note that the neutrinospheric temperatures, neutrino opacities, and EOS employed here
differ from what was used in those studies, hence numerical values are expected to differ.

The calculation of this limiting luminosity in a consistent manner with the microphysics, initial
conditions, and boundary conditions employed in our simulations allows direct testing of the
hypothesis that the critical stability threshold for explosion is given by the Burrows-Goshy limit.
We find that in all cases, the measured thresholds for both type of mode lie below this limiting value. 
When the neutrinospheric radius is held constant, the instability thresholds approach (but never
equal) the Burrows-Goshy limit for increasing mass accretion rate.
In the sequence that relates the neutrinospheric radius to the neutrino luminosity via
the black body relation (Figure~\ref{f:BG93_data}d), the instability thresholds have a 
nearly constant separation in luminosity over the entire range of accretion rates investigated.

The sequence with $R_\nu = 20$~km yielded unexpected results for 
$\dot{M} \ge 0.5M_\sun$~s$^{-1}$, however. In this case there is neither instability nor runaway
for all luminosities up to and including the limiting value, independent of numerical resolution.
Our approximate instability criteria fail to predict this behavior, showing that additional
constraints play a role in determining instability. Also, the fact
that no explosion is found at the Burrows-Goshy limit shows that this luminosity is 
not an independent tracer of stability either. Increasing the neutrino luminosity above
the limiting value (at constant shock radius) by $\sim 10\%$ causes the shock to readjust to a new equilibrium
position, which is also stable. It is worth noting however that the shock radius is $R_s \le 61$~km
for the non-exploding segment of this sequence. This is an unrealistically low value, with advection
times of the order of $\sim 3$~ms, a factor of a few from the dynamical time. It is likely that
the fixed value of $r_{\rm vf}$ used to set the upstream velocity (\S\ref{s:initial_conditions})
leads to these extreme conditions.

In all other sequences, the approximate instability criteria found in \S\ref{s:approximate} provide a 
good description of the measured values over a wide region of parameter space, with agreement better 
than $5\%$ in $L_{\nu_e}$. They correctly capture the disappearance 
of the oscillatory mode in the $R_\nu = 30$~km sequence at high accretion rates, 
where only a non-oscillatory mode is measured. This phenomenon occurs because at large 
accretion rates, the shock radius becomes increasingly smaller. The size of the resulting gain region is such
that the advection time through it never exceeds $t_{\rm adv-e}$ before the heating time drops below $t_{\rm adv-g}$
due to the increasing luminosity. The deviation of the measured oscillatory threshold from the approximate
criterion at high accretion rate in the $R_\nu=40$~km sequence is due to $r_e$ not being a good tracer of the 
point where pressure and density perturbations are 180 degrees out-of-phase (\S\ref{s:approximate}).

Our results extend the findings of \citet{yamasaki07}, who obtained $\ell=0$ instability thresholds
below the Burrows-Goshy limit, to a wider region of parameter space. Our results are
quantitatively different from theirs, partly due to the different neutrinospheric temperatures that they employed
($T_{\nu_e} = T_{\bar\nu_e} = 4.5$~MeV), their use of the pure absorption 
coefficient for computation of the optical depth (see \S\ref{s:initial_conditions} and \S\ref{s:other}), and
their inclusion of self-gravity. 
Using the same parameters as \citet{yamasaki07}, we obtain a limiting luminosity of
$L_{\rm BG,\nu_e} = 7.36\times 10^{52}$~erg~s$^{-1}$ for $\dot{M}=1M_\sun$~s$^{-1}$,
which differs by a factor of nearly two, and is closer to what
was found originally by \citet{burrows93}. 
The corresponding approximate instability criteria for oscillatory and
non-oscillatory modes are $6.41\times \times 10^{52}$~erg~s$^{-1}$ and $7.04\times \times 10^{52}$~erg~s$^{-1}$.
These values are lower than our limiting luminosity by 13\% and 5\%, respectively.

\subsection{Other Parameter Dependencies}
\label{s:other}

Given the parametric character of this study, certain choices had to be
made in order to construct a background flow that is as realistic as possible. 
Here we explore how our results depend on numerical resolution, boundary conditions, parameters
that determine the upstream flow, and the suppression of source terms near the neutrinosphere.

\begin{figure}
\includegraphics*[width=\columnwidth]{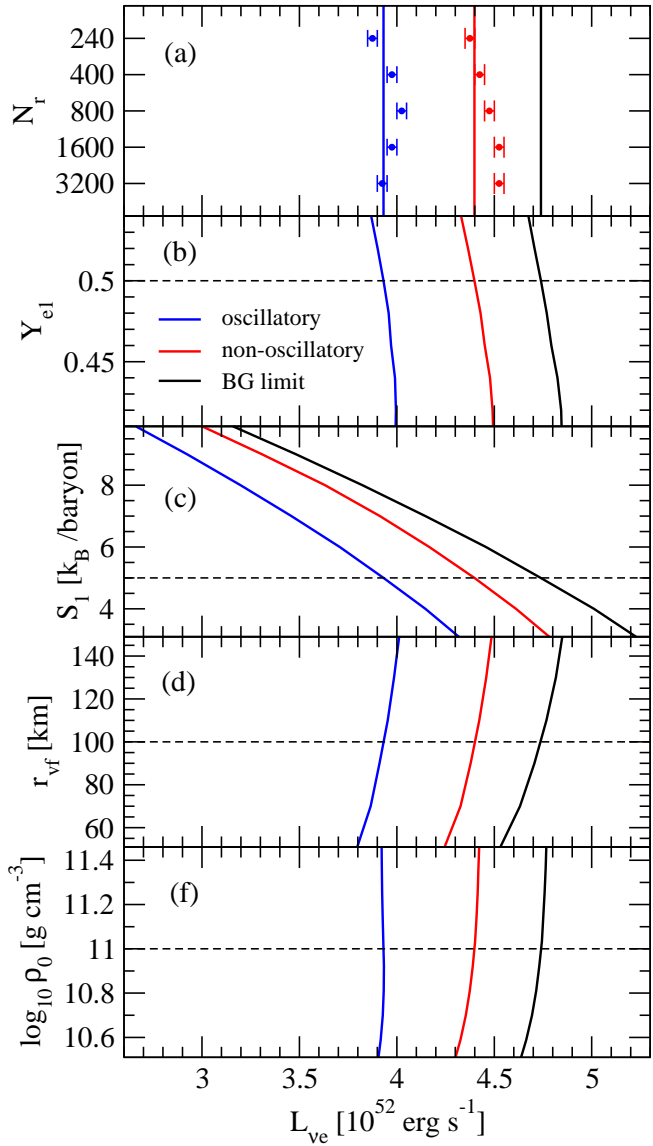}
\caption{Dependence of the approximate instability thresholds and Burrows-Goshy limit on various
parameters, taking our fiducial sequence as baseline. Solid lines denote the approximate instability condition for oscillatory
(blue, eq.~[\ref{eq:instability_osc}]) and non-oscillatory modes (red, eq.~[\ref{eq:instability_nonosc}]), respectively, 
while the black solid line denotes the Burrows-Goshy limit. Panel (a) shows simulation results (as data points)
as a function of radial resolution in a logarithmic grid from $30$~km to $1000$~km. Other panels
show the dependence of these critical points on the upstream electron fraction (b), upstream entropy (c), radius at which
upstream velocity is set to the free-fall speed (d), and density cutoff for source terms (d, eq.~[\ref{eq:suppression}]).
Parameters of our fiducial sequence are denoted by a black dashed line.}
\label{f:Lnu_sim_parameters}
\end{figure}

Figure~\ref{f:Lnu_sim_parameters}a shows the approximate instability criteria and Burrows-Goshy
limit for our fiducial sequence, together with instability thresholds measured from simulations 
at different resolutions. The grid
is chosen logarithmically spaced, with a ratio of spacing between adjacent cells
$\zeta = (R_{\rm max}/R_\nu)^{1/N_r}$, and $\Delta r_{\rm min} = R_\nu(\zeta-1)$,
with $\Delta r_{\rm min}$ the cell adjacent to the inner boundary (Appendix~\ref{s:grid_test}).
Oscillatory modes asymptote to the theoretical threshold for increasing resolution,
albeit convergence is non-monotonic. Non-oscillatory modes converge monotonically
to a slightly different value, indicating that the instability criterion is indeed approximate.

The results shown in Figure~\ref{f:Lnu_sim_parameters} also indicate that previous hydrodynamic studies
that used similar methods to solve the hydrodynamic equations \citep{murphy08,nordhaus10a,hanke11}
have enough resolution to capture the critical stability thresholds correctly.
However, we have found in our models that there are noticeable
differences in the growth rates of oscillatory modes as a function of resolution. For the model
with $L_{\nu_e,52}=4.1$ in our fiducial sequence, the time delay to explosion varies by $\sim 100$~ms 
when going from $N_r=240$ to $N_r=3200$.
True convergence, in the sense that the fractional deviations in the shock radius evolution 
are much less than unity for models at different resolutions, is only achieved for $N_r \ge 1600$ 
with our fiducial parameters. It is not straightforward to prescribe a definite
resolution required for convergence given a hydrodynamic method, however, as the implementation
of the microphysics (tabular in our case) also influences this number. A set of standard hydrodynamic
tests tailored specifically for the supernova problem would be helpful in assessing the 
reliability of results associated with a given implementation.

\begin{figure}
\includegraphics*[width=\columnwidth]{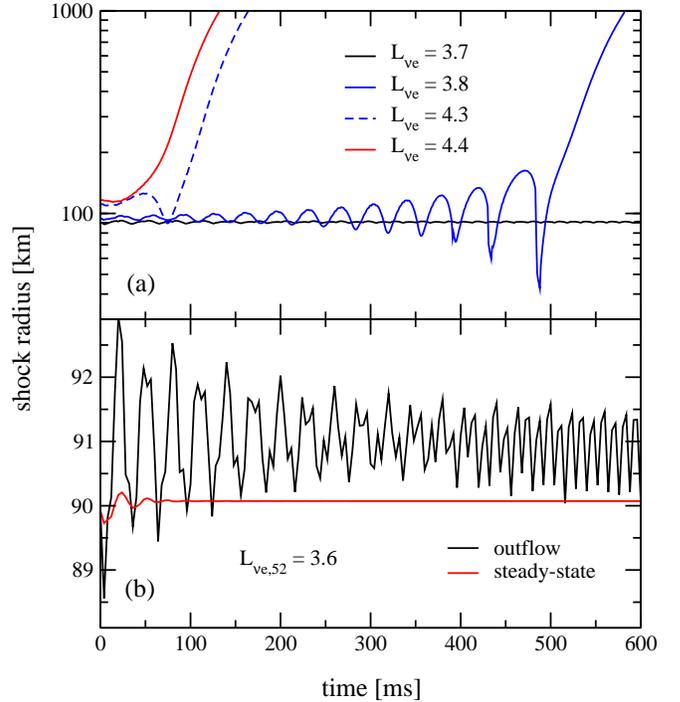}
\caption{Evolution of the shock radius in selected models of our fiducial sequence using different
prescriptions for the inner boundary condition. 
Panel (a) shows models with different neutrino luminosities using the outflow boundary condition 
(eqns.~[\ref{eq:outbnd2_dens}]-[\ref{eq:outbnd2_velr}]). The curves shown bracket the instability 
threshold for oscillatory (black and solid blue) and non-oscillatory (dashed blue and red) modes. 
Except for the boundary condition, parameters are identical to those in Figure~\ref{f:shock_radius_1d}.
Panel (b) compares two runs with the same parameters except the inner boundary condition. Both models are stable 
and do not explode (note the vertical scale).}
\label{f:rshock_outbnd2}
\end{figure}

To test the influence of the steady-state boundary condition on our results, we have evolved
a sequence with a different inner boundary condition, which allows an arbitrary amount
of mass to leave the domain, while providing enough pressure support to prevent the shocked
envelope from collapsing (\S\ref{s:time_dependent}). Figure~\ref{f:rshock_outbnd2}a shows that the system 
still goes through the same stability phases as the neutrino luminosity is increased,
with a small quantitative difference $\lesssim 5\%$ in the critical stability points relative 
to our fiducial sequence.
We thus conclude that the two types of instability do not depend on the
fluxes of mass, momentum, and energy leaving the domain being fixed. The quantitative
differences can be explained by the fact that the outflow boundary condition acts as a persistent
source of waves, keeping shock oscillations at some non-zero amplitude.
This effect is shown in Figure~\ref{f:rshock_outbnd2}b, which compares the shock radius evolution for 
two stable runs with identical parameters except for the boundary condition.

\begin{figure}
\includegraphics*[width=\columnwidth]{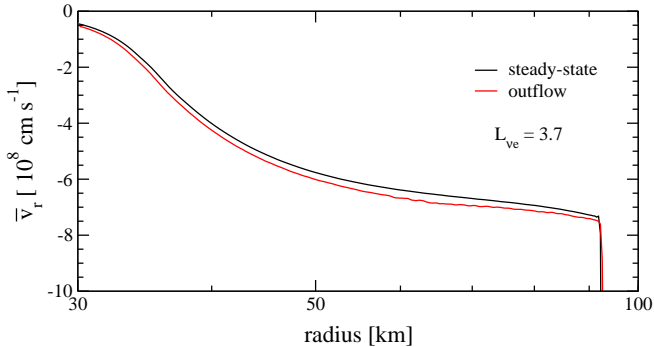}
\caption{Time averaged radial velocity as a function of radius for two stable models in our fiducial
sequence with the same parameters except the inner boundary condition. The time average is taken after the
models have settled into a steady- or quasi-steady state.}
\label{f:velr_outbnd2}
\end{figure}

This small quantitative difference, caused by a persistent wave source, is interesting as 
it provides an illustration of what we expect will occur in the multidimensional case. 
The critical stability thresholds are modified in the presence of this additional wave generation, as
comparison of Figures~\ref{f:shock_radius_1d} and \ref{f:rshock_outbnd2} show.
Taking the time-average of two stable runs with differing boundary conditions after they have settled
into steady- or quasi-steady-state, shows that the mean velocity profiles differ (Figure~\ref{f:velr_outbnd2}). 
The magnitude of the change in the threshold luminosities is consistent with the magnitude and direction 
of the changes in the approximate instability criteria, using the mean advection times rather than the
unperturbed ones. A stronger source of wave energy (e.g., convection) will affect the
mean advection time in stronger way, and critical stability points will move accordingly.

Figures~\ref{f:Lnu_sim_parameters}b-e illustrate the sensitivity of the approximate instability thresholds 
and the Burrows-Goshy limit to the parameter choices made in \S\ref{s:initial_conditions}.
Shown is the dependence on the upstream electron fraction, upstream entropy, radius at which
the upstream flow has free-fall speed, and cutoff density for neutrino source terms (eq.~[\ref{eq:suppression}]).
There is a somewhat sensitive dependence on the upstream entropy, which in a realistic situations will be set
by the entropy in the progenitor core. Note that larger luminosities are needed to cause stalled shocks to become
unstable when the progenitor entropy is smaller. Nonetheless, the relation between the approximate thresholds and 
the Burrows-Goshy limit persists without qualitative changes. All other parameters produce changes smaller than 
$\sim 5\%$ over a large range of values.

\begin{figure}
\includegraphics*[width=\columnwidth]{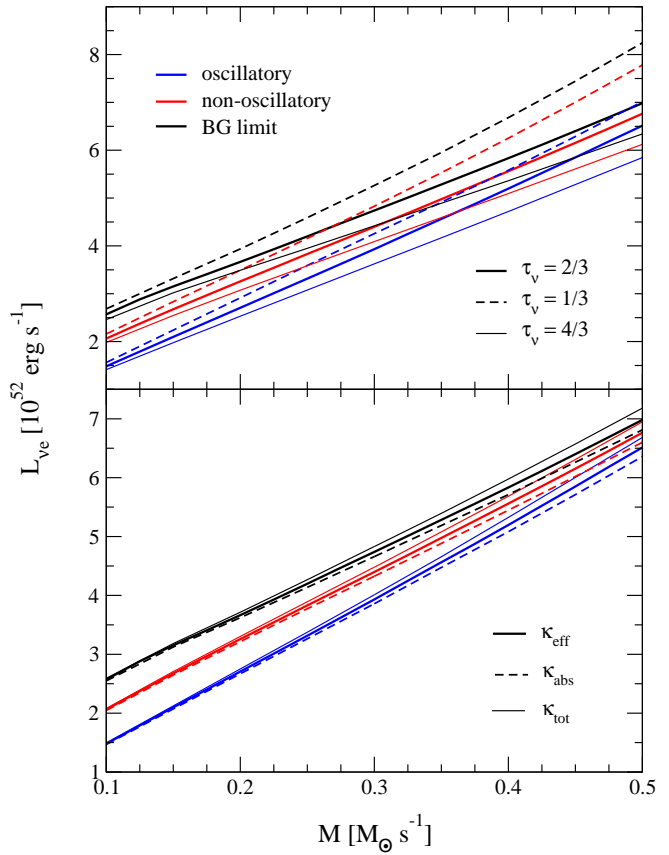}
\caption{Approximate instability thresholds and Burrows-Goshy limit as a function of mass accretion rate
for different optical depths (top) and different prescriptions for the neutrino opacity (bottom). The 
three curves remain separated.}
\label{f:Lbg_taukappa}
\end{figure}

We have also examined the robustness of the hierarchy between instability thresholds and Burrows-Goshy limit
when the quantities used to calculate the latter are changed. Figure~\ref{f:Lbg_taukappa}a shows the effect
of increasing or decreasing the neutrinospheric optical depth (eq.~[\ref{eq:optical_depth_def}]) by a factor of two.
Neither of these changes cause the Burrows-Goshy limit or instability thresholds to cross each other, for
a given optical depth.
This result can also be seen from the non-overlapping curves in Figure~\ref{f:optical_depth_curve} for fixed accretion rate.
Interestingly, the exact optical depth chosen seems to matter the least when going to low mass accretion rates.

The dependence on the choice of neutrino opacity is shown in Figure~\ref{f:Lbg_taukappa}b. Here we compare 
the effects of using the pure absorption, effective, or total (absorption plus scattering) opacity (\S\ref{s:initial_conditions})
in equation~(\ref{eq:optical_depth_def}). Again, the three curves maintain their relative hierarchy and do not
cross for a given choice of opacity.

\vspace{0.5in}

\subsection{Relation between the Burrows-Goshy Limit and the Instability Thresholds}

We have found that the difference between the Burrows-Goshy limit and the radial
instability thresholds is finite and measurable, and is independent of parameter choices. 
However, the question remains as of why is this limiting luminosity 
always close to the threshold for non-oscillatory instability in the parameter region relevant
to core-collapse supernovae.

In a more extended parameter space study of steady-state configurations, \citet{pejcha2011} 
found that at the Burrows-Goshy limit, the ratio between $t_{\rm adv-g}$
and the global heating timescale (without gravity) is always close to unity.
Ultimately, this proximity must depend on the characteristic magnitude of the neutrino opacities, 
the gravitational binding energy around a forming neutron star, and the fact that the flow 
is supported against gravity by the thermodynamic pressure of the gas (e.g., not centrifugal forces). 
In other words, it could be just a result of dimensionality.
A possible way to test this hypothesis is to explore the behavior of the system at lower mass accretion rates, 
where the expected trend is an increasing deviation. 

\vspace{1in}

\section{Summary and Discussion}
\label{s:summary}

We have investigated the transition to runaway expansion of a stalled core-collapse 
supernova shock in spherical symmetry, when the parameters that describe the 
accretion flow are varied systematically. A realistic equation of state and
weak interactions were employed to perform time-dependent simulations with FLASH3.2.
Starting from steady-state solutions to the hydrodynamic equations, we evolve 
sequences of time-dependent models with increasing neutrino luminosity, and analyze the 
hydrodynamic processes that mediate the transition from accretion to runaway expansion. 
Our findings can be summarized as follows:
\newline

\noindent 1. -- The onset of radial instability is a sufficient condition for runaway expansion
		when heating by the accretion luminosity is ignored and 
        	neutrinospheric parameters remain constant in time. 
                Radial instability can manifest itself via oscillatory and non-oscillatory modes,
		as found in the linear stability analysis of \citet{yamasaki07} and 
		numerous time-dependent hydrodynamic studies.
	 	These modes are non-adiabatic, as they involve changes in the heat content of
		the fluid (\S\ref{s:linear}).
		\newline

\noindent 2. -- For both types of modes, transition to runaway expansion occurs after a portion
 	        of the fluid in the gain region achieves positive energy.
		This coincides with the fluid just below the shock achieving positive velocity,
		starting the phase of runaway expansion (\S\ref{s:nonlinear}; Figures~\ref{f:etot_velx_osc} 
		and \ref{f:etot_velx_nonosc}). 
		\newline

\noindent 3. -- The only significant source of dissipation in our models is the energy loss
		from shock motion (eq.~[\ref{eq:Edot_s}]; Figures~\ref{f:dEdt_oscillatory} and 
		\ref{f:dEdt_nonoscillatory}), which provides damping on expansion.
		Nuclear recombination and a steady source of neutrino heating combine to ensure
		that this term never extinguishes the runaway or saturate the linear instability. 
		In a more realistic context,
		the dominant dissipative processes are the
		decrease of the core neutrino
		luminosity with time, and self consistent heating by the accretion luminosity. 
		The latter is expected to provide a negative feedback during expansion due
		to the decrease of the mass accretion rate (e.g., \citealt{janka96}).
		The contraction of the protoneutron star is a somewhat smaller correction,
		of the order of $\sim 10\%$.
		None of these effects is included in this study (\S\ref{s:saturation}). 
		\newline
         
\noindent 4. -- We have found approximate instability criteria for oscillatory and non-oscillatory modes
		that correctly describe the behavior of the system over a wide region of parameter
		space, with a precision better than $5\%$ in neutrino luminosity (\S\ref{s:approximate}; 
		Figure~\ref{f:BG93_data}). For oscillatory modes,
	 	instability arises when the advection time over the gain region becomes longer than the
		advection time from the gain radius to the point where the pressure and density perturbations
		are 180 degrees out of phase (Figures~\ref{f:spacetime} and \ref{f:instability_criteria}). 
		In our implementation, this point lies at a position in the background flow where the cooling has decreased by
		an e-folding from its peak value. This equality of advection times is equivalent to an
		equality of masses of the respective regions when the accretion rate is constant.
		We have not been able to identify a conclusive physical reason for why this condition on the masses
		or advection times triggers an oscillatory instability. We surmise that a relation might exist 
		between the fraction of the oscillation cycle that the fluid gains energy and the phase lag 
		criterion in stellar pulsation theory (e.g., \citealt{cox74}).
		\newline
	        
\noindent 4. -- Non-oscillatory modes become unstable when the advection time through the
		gain region becomes longer than the integrated total energy divided by the
		integrated net heating in the gain region. This condition means that heating
		is effective at increasing the thermal energy while the fluid transits the
		gain region, increasing the pressure and causing the shock to adjust to a
		new equilibrium position \citep{janka98}. This global criterion has been used
		by a number of previous studies as an explosion diagnostic \citep{buras06b,murphy08,marek09}.
		\newline

\noindent 5. -- The instability thresholds are in general different from the limiting luminosity for
		the steady-state system (\citealt{burrows93}, \S\ref{s:limiting_luminosity}).
		For constant neutrinospheric radius and increasing
		accretion rates, the thresholds asymptote to the limiting luminosity, but never
		coincide with it (Figure~\ref{f:BG93_data}). This separation does not depend on how
		this limiting luminosity is calculated (Figure~\ref{f:Lbg_taukappa}),
		or on specific parameter choices (Figure~\ref{f:Lnu_sim_parameters}).
		\newline

\noindent 6. -- The existence of a limiting luminosity for steady-state solutions is a direct consequence 
                of requiring the optical depth
		between the neutrinosphere and the shock to have a fixed value (Figure~\ref{f:optical_depth_curve}). For fixed
		upstream conditions and increasing neutrino luminosity, the entropy increases
		and hence the radius at which the pressure transitions from being dominated
		by relativistic particles to being dominated by non-relativistic nucleons 
		moves inward (Figure~\ref{f:profiles_taupeak}). This causes the density profile to soften at 
		fixed radius, resulting in a lower
		density at the neutrinosphere. Above a certain limit, there is not enough
		mass to provide sufficient neutrino optical depth to satisfy the closure
		relation, and no steady-state solution is possible. This result is equivalent to that
		of \citet{pejcha2011}, differing only in which boundary conditions are assumed to
		be fulfilled.
		\newline

\noindent 7. -- We find neither instability nor explosion in our sequence with constant $R_\nu = 20$~km
	        for $\dot{M} \ge 0.5M_\sun$~s$^{-1}$ and luminosities up to and including the 
		limit luminosity (Figure~\ref{f:BG93_data}b). At the lower end of these mass accretion rates,
		the shock radius is $R_s \simeq 60$~km, and decreases for larger accretion rates. We did not
		find an explanation for this result within our framework.
		Regardless of the reason, however, it
	 	shows that the transition to runaway expansion does not necessarily occur
		at the limiting luminosity. It also shows that the approximate instability criteria
		derived here are incomplete, as an additional constraint is likely to determine the stability
		of the flow.
		\newline

Our results show that the \citet{burrows93} conjecture, relating the transition to explosion to a global
instability of the shocked envelope, is correct within a restricted set of assumptions. The critical
stability surface, however, is not given by the limiting luminosity of the steady-state configuration.
This limiting luminosity remains nonetheless close to the instability 
threshold over a large section of the parameter space relevant to core-collapse supernovae, a result that we have
not accounted for and which deserves further investigation.

\citet{murphy08} and \citet{FT09b} have also found oscillations around the transition to explosion. However,
the width in neutrino luminosity separating exploding from non-exploding configurations is larger in 
those studies. In the case of \citet{FT09b}, this can be attributed to the differences in the employed microphysics
relative to the present study. Large amplitude oscillations are excited in the simulations of \citet{murphy08} 
when the Si/O composition interface is accreted through the shock. Stable models damp oscillations, and unstable
modes explode, with neutrally stable oscillations confined to a range $\Delta L_{\nu_e}/L_{\nu_e}\simeq 4\%$.
This range is wider than that found here, presumably because \citet{murphy08} include the contraction of the
protoneutron star. Our results are in good agreement with those of \citet{ohnishi06}, who used essentially the same 
physical assumptions as we do.

The main result of this paper, point (1) above, does not apply to all modes of higher dimensionality. 
Non-radial oscillatory instability of the shock (the SASI) does not necessarily lead
to explosion when the same set of assumptions adopted in this study are employed \citep{ohnishi06}. 
Given our results and those of \citet{yamasaki07},
the question then arises as to whether a multi-dimensional explosion can be thought of
as the excitation of an unstable radial-oscillatory, and/or non-oscillatory mode (radial or otherwise)
by the action of turbulent stresses from the SASI and convection.
In this case, the  excited mode would arise from a background state that differs from the laminar steady-state solution
by the presence of a turbulent pressure term in the momentum equation and a convective
flux term in the energy equation. The excited mode would also become unstable at a lower neutrino
luminosity than in the spherically symmetric flow.

According to \citet{yamasaki07}, the non-radial-non-oscillatory modes that have the lowest threshold
luminosity for instability have Legendre indices $\ell\sim 6$.
These modes are likely to be associated with convection in the gain region, as they have no unstable oscillatory
counterpart. The results of \citet{ohnishi06} and \citet{iwakami08} show that small-scale convection in itself does 
not lead to runaway expansion. 
In contrast, an $\ell=1$ or $\ell=2$ non-oscillatory mode, such as that envisioned by \citet{thompson00}, is likely
to be behind unipolar or bipolar explosions seen in axisymmetric simulations (e.g., \citealt{scheck06}).
The results of \citet{yamasaki07} show that the threshold luminosities of these large scale
non-oscillatory modes are larger than that of $\ell \ge 3$ modes, and very close to that
of the spherical oscillatory mode. 

Which of these modes is excited first in a multidimensional context will
obviously depend on the nature of the new background flow. The exploding 
three-dimensional models of \citet{nordhaus10a} and \citet{hanke11} 
lack a dominant oscillatory mode in their transition to explosion, even though
oscillations in the average shock position are still visible (particularly for the
$L_{\nu_e}=9\times 10^{51}$~erg~s$^{-1}$ model in the 11.2$M_\sun$ progenitor
sequence of \citealt{hanke11}). In contrast, two-dimensional models
have a noticeable radial oscillatory component, with successive dips in the average
shock radius before and during runaway expansion. Yet the amplitude of these oscillations
is smaller than in the spherically symmetric case, suggesting that more than one mode
is likely to be involved.
 
Support for this interpretation of multidimensional explosions can be found in the 
results of \citet{buras06b} and \citet{marek09}, who
use the same definition of $t_{\rm heat-g}$ as we do here. They find that the ratio $t_{\rm adv-g}/t_{\rm heat-g}$
exceeding unity corresponds roughly to the time when the shock begins its expansion towards explosion in
two-dimensional runs. Obviously
all the saturation mechanisms discussed in \S\ref{s:saturation} are present in those simulations, so 
analogies need to be made cautiously. A similar result is obtained by \citet{murphy08}, with
a heating time defined without gravity and kinetic energy.

An attempt to include the effects of convection in the steady-state solution was made by \citet{yamasaki06},
setting the convective flux to a value that yielded a flat entropy gradient. With this maximally efficient
convection, they found that the Burrows-Goshy limit can decrease by several tens of percent
relative to the laminar case. This would entail a corresponding decrease in the approximate instability
thresholds found in this paper.
A more careful treatment of the turbulent transport terms, along the
lines of \citet{murphy11}, could be used to construct quasi-steady-state flows and to perform 
a linear analysis like that of \citet{yamasaki07}, helping to elucidate these questions with
semianalytic tools.

The extension of our model to two- and three spatial dimensions will be discussed in a companion paper.

\acknowledgements
I thank Aristotle Socrates, Thomas Janka, Thierry Foglizzo, Jeremiah Murphy, Adam Burrows,
Jason Nordhaus, and Boaz Katz for stimulating discussions.
I also thank Evan O'Connor for help with the EOS implementation.
Comments on the manuscript by an anonymous referee provided useful feedback and 
helped to improve the presentation.
The author is supported by NASA through Einstein Postdoctoral Fellowship
grant number PF-00062, awarded by the Chandra X-ray Center, which is operated
by the Smithsonian Astrophysical Observatory for NASA under contract NAS8-03060.
This research was supported in part by the National Science Foundation through 
TeraGrid resources \citep{catlett07}. Computations were performed at the NCSA Abe,
LONI Queen Bee, and IAS Aurora clusters.
The software used in this work was in part developed by the DOE-supported ASC / 
Alliance Center for Astrophysical Thermonuclear Flashes at the University of Chicago.
\newline

\appendix

\section{Weak Interaction Rates}
\label{s:weak_rates}

Here we describe the calculation of weak interaction source
terms entering the energy and lepton conservation equations 
(eqns.~[\ref{eq:energy_conservation}] and [\ref{eq:lepton_conservation}], respectively).
We only consider electron-type neutrinos and antineutrinos, as other species do not
exchange energy with matter outside the neutrinosphere.
Our implementation largely resembles that of \citet{ohnishi06}, with some minor modifications.

For both species, we assume a free-streaming neutrino flux radiated isotropically 
from neutrinospheres located at the same radius $R_\nu$. The neutrino distributions are
approximated by a Fermi-Dirac function with zero chemical potential
\begin{equation}
\label{eq:fnu_definition}
f_{\nu_i}(\epsilon,T_{\nu_i},r,\cos\theta_k)  =  N_{\nu_i}\,F_{\rm FD}(\epsilon,T_{\nu_i},0)\,
\Theta\left[\cos\theta_k - \cos\theta_\nu(r)\right],
\end{equation}
where $\nu_i = \{\nu_e,\bar\nu_e\}$,
\begin{equation}
F_{\rm FD}(\epsilon,T,\mu) = \frac{1}{\exp{(\epsilon-\mu)/kT} + 1}
\end{equation}
is the Fermi-Dirac distribution with energy $\epsilon$ and chemical potential $\mu$.
The normalization factor
\begin{equation}
\label{eq:fnu_normalization}
N_{\nu_i}  =  \frac{L_{\nu_i}}{(7/16)\, 4\pi R_\nu^2\, \sigma_{\rm SB} T_{\nu_i}^4}\\
\end{equation}
is such that the neutrino luminosity $L_{\nu_i}$ and spectral temperature $T_{\nu_i}$ are
independent parameters. The step function $\Theta$ contains the angular dependence
of the radiation field, which propagates up to an angle
\begin{equation}
\theta_k < \theta_\nu(r)  =  \arccos\left[1 - \left(\frac{R_\nu}{r}\right)^2\right]^{1/2}
\end{equation}
from the radial direction. In equation~(\ref{eq:fnu_normalization}), $\sigma_{\rm SB}$
is the Stefan-Boltzmann constant.

To approximate the results of radiation-hydrodynamic simulations
with our simplified assumptions, we set $L_{\nu_e} = L_{\bar{\nu}_e}$  but specify
different temperatures, $T_{\nu_e} = 4$~MeV and $T_{\bar{\nu}_e}=6$~MeV, reflecting 
the fact that the photospheres are at slightly different locations, but with a spectrum
such that luminosities are roughly comparable \citep{janka95,janka01}. We ignore however
the correction due to the spatial difference between the electron neutrino- and antineutrinospheric
radius, which is of the order of 10\%. This difference is absorbed by the normalization 
factor in equation~(\ref{eq:fnu_normalization}). 

The evolution of $Y_e$ above the neutrinosphere is determined by the net production rate
of electron and positrons $\Gamma_{e-}$ and $\Gamma_{e+}$,
respectively (eq.~[\ref{eq:lepton_conservation}]), due to neutrino absorption and emission. 
Following \citet{bruenn85}, these are given by
\begin{eqnarray}
\Gamma_{\rm net} & = & \Gamma_{\rm e^{-}} - \Gamma_{\rm e^{+}}\\
\label{eq:gamma_nem}
\Gamma_{e-} & = & \frac{2\pi m_{\rm n} c}{(hc)^3\rho}\int \totd\cos\theta_k \int \epsilon^2 \totd\epsilon
                    \left[ \kappa_{\nu_e} f_{\nu_e} - j_{\nu_e}\left(1-f_{\nu_e} \right) \right]\\
\label{eq:gamma_nep}
\Gamma_{e+} & = &\frac{2\pi m_{\rm n} c}{(hc)^3\rho}\int \totd\cos\theta_k \int \epsilon^2 \totd\epsilon
                    \left[ \kappa_{\bar\nu_e} f_{\bar\nu_e} - j_{\bar\nu_e}\left(1-f_{\bar\nu_e} \right)\right],
\end{eqnarray}
where $j_{\nu_i}$ and $\kappa_{\nu_i}$ are the emissivity and absorption coefficient, respectively, associated with
electron-type neutrinos or antineutrinos (as subscripted), and the integrals are performed over all propagation angles and
positive energies. Expressions for these coefficients are obtained by \citet{bruenn85}
assuming detailed balance, matter in nuclear statistical equilibrium, and non-relativistic nucleons, whose
recoil is neglected. In addition, we ignore here the nucleon phase space blocking factors, which deviate
only slightly from unity at densities $\lesssim 10^{11}$~g~cm$^{-3}$ \citep{bruenn85}, yielding
\begin{eqnarray}
j_{\nu_e}(\epsilon,T,\mu_e,n_p) & = & \frac{\tilde{G}_{\rm F}^2}{\pi}\left(g_{\rm V}^2 + 3g_{\rm A}^2\right)\, n_p
                                    F_{\rm FD}(\epsilon,T,\mu_e)\left[\epsilon + \Delta_m \right]^2
                                    \times\left[1 - \frac{m_e^2 c^4}{(\epsilon + \Delta_m)^2}\right]^{1/2}\\
j_{\bar{\nu_e}}(\epsilon,T,\mu_e,n_p) & = & \frac{\tilde{G}_{\rm F}^2}{\pi}\left(g_{\rm V}^2 + 3g_{\rm A}^2\right)\, n_n
                                    F_{\rm FD}(\epsilon-\Delta_m,T,-\mu_e)\left[\epsilon - \Delta_m \right]^2
                                    \times\left[1 - \frac{m_e^2 c^4}{(\epsilon - \Delta_m)^2}\right]^{1/2}\nonumber\\
				& &	\Theta(\epsilon-\Delta_m - m_e c^2)\\
\kappa_{\nu_e}(\epsilon,T,\mu_e,n_p) & = & \frac{\tilde{G}_{\rm F}^2}{\pi}\left(g_{\rm V}^2 + 3g_{\rm A}^2\right)\, n_n
                                    \left[1-F_{\rm FD}(\epsilon,T,\mu_e)\right]\left[\epsilon + \Delta_m \right]^2
                                    \times\left[1 - \frac{m_e^2 c^4}{(\epsilon + \Delta_m)^2}\right]^{1/2}\\
\kappa_{\bar{\nu_e}}(\epsilon,T,\mu_e,n_p) & = & \frac{\tilde{G}_{\rm F}^2}{\pi}\left(g_{\rm V}^2 + 3g_{\rm A}^2\right)\, n_p
                                    \left[1-F_{\rm FD}(\epsilon-\Delta_m,T,-\mu_e)\right]\left[\epsilon - \Delta_m \right]^2
                                    \times\left[1 - \frac{m_e^2 c^4}{(\epsilon - \Delta_m)^2}\right]^{1/2}\nonumber\\
				& &	\Theta(\epsilon-\Delta_m - m_e c^2),
\end{eqnarray}
where $\mu_e$ is the chemical potential of electrons, $n_n$ and $n_p$ the number density of free neutrons and protons, respectively,
$\tilde{G}_{\rm F} = G_F/(\hbar c)^3$ the Fermi constant, $g_{\rm V}$ and $g_{\rm A}$ the vector
and axial coupling constants, respectively, and $\Delta_m = (m_n - m_p) c^2$ the difference between the
rest mass energy of neutrons and protons.

The rates of energy exchange between neutrinos and matter are similarly found as a function
of $f_{\nu_i}$, $j_{\nu_i}$, and $\kappa_{\nu_i}$. The heating and cooling rates per unit volume
are given respectively by
\begin{eqnarray}
\label{eq:L_H_def}
\mathscr{L}_H                   & = & \frac{\rho}{m_n}\left( Q_{\nu_e}^{+} + Q_{\bar\nu_e}^{+} \right)\\
\label{eq:L_C_def}
\mathscr{L}_C                   & = & \frac{\rho}{m_n}\left( Q_{\nu_e}^{-} + Q_{\bar\nu_e}^{-} \right),
\end{eqnarray}
where $Q_{\nu_i}^{\pm}$ are rates per baryon of heating (+ superscript) and cooling (- superscript) 
due to electron-type neutrinos ($\nu_e$ subscript) and antineutrinos ($\bar\nu_e$ subscript). These rates
are given by 
\begin{eqnarray}
\label{eq:heating_nem}
Q_{\nu_e}^+     & = &  \frac{2\pi m_{\rm n} c}{(hc)^3\rho}\int \totd\,\cos\theta_k \int \epsilon^3 \totd\,\epsilon
                       \left[ j_{\nu_e} + \kappa_{\nu_e}\right]f_{\nu_e}\\
Q_{\bar\nu_e}^+ & = &  \frac{2\pi m_{\rm n} c}{(hc)^3\rho}\int \totd\,\cos\theta_k \int \epsilon^3 \totd\,\epsilon
                       \left[ j_{\bar\nu_e} + \kappa_{\bar\nu_e}\right]f_{\bar\nu_e}\\
\label{eq:cooling_nem}
Q_{\nu_e}^-     & = &  \frac{4\pi m_{\rm n} c}{(hc)^3\rho}\int \epsilon^3 \totd\,\epsilon \, j_{\nu_e}\\
\label{eq:cooling_nep}
Q_{\bar\nu_e}^- & = &  \frac{4\pi m_{\rm n} c}{(hc)^3\rho}\int \epsilon^3 \totd\,\epsilon \, j_{\bar\nu_e}.
\end{eqnarray} 

\begin{figure*}
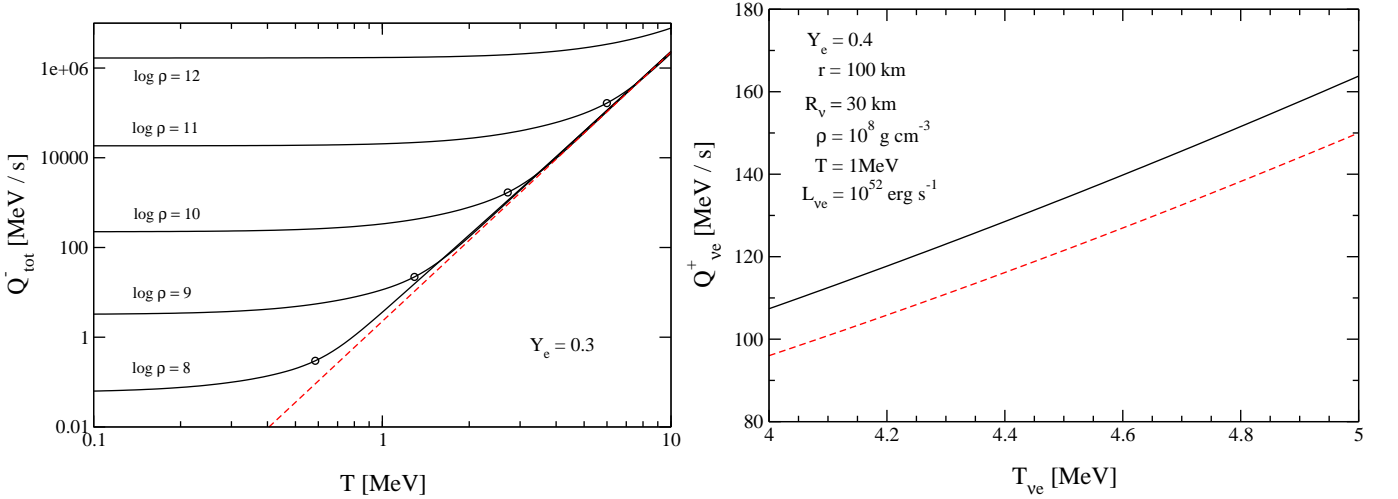

\includegraphics*[width=0.5\textwidth]{f16a.eps}
\includegraphics*[width=0.5\textwidth]{f16b.eps}
\label{f:cooling_heating_app}
\caption{\emph{Left}: Total cooling rate per baryon $Q^-_{\rm tot} = 
Q^-_{e-}+Q^-_{e+}$ (eqns.~[\ref{eq:cooling_nem}]-[\ref{eq:cooling_nep}]) 
as a function of temperature, for fixed electron fraction $Y_e = 0.3$, and assuming $X_p+X_n=1$.
Different solid lines correspond to different densities (in g~cm$^{-3}$), with circles
marking the approximate boundary between degenerate and non-degenerate regimes, at
$\mu_e = \pi kT$. The red dashed line shows the approximate expression $Q^-_{\rm tot} \simeq 145 (kT/2\textrm{MeV})^6$~MeV~s$^{-1}$,
obtained assuming non-degenerate electrons \citep{janka01}.
\emph{Right:} Heating due to absorption of electron-type neutrinos (eq.~[\ref{eq:heating_nem}]) as a function of the temperature of the
electron neutrinosphere $T_{\nu_e}$, for parameters as shown in the panel. For reference, the red-line shows
the approximate expression $Q^+_{\nu e} \simeq 160 L_{\nu,52} r_7^{-2} (1-Y_e) (kT_{\nu_e}/4\textrm{MeV})^2$~MeV~s$^{-1}$
from \citet{janka01}. 
}
\end{figure*}

Numerical calculation of equations~(\ref{eq:gamma_nem})-(\ref{eq:gamma_nep})
and (\ref{eq:heating_nem})-(\ref{eq:cooling_nep}) involves tabulating Fermi-Dirac integrals as a function of $T$, $\mu_e$,
and $T_{\nu_i}$, with other dependencies entering as global scaling factors.
The left panel of Figure~\ref{f:cooling_heating_app} shows the total cooling per baryon 
$Q^-_{\rm tot} = Q^-_{\nu_e}+Q^-_{\bar\nu_e}$ as a function of temperature, for fixed $Y_e$ and different
densities. 
In the degenerate regime $kT < \pi\mu_e$,
the cooling rates are nearly independent of temperature, whereas for higher temperatures they approach the
asymptotic expression $Q^-_{\rm tot} \simeq 145 (kT/2\textrm{MeV})^6$ from \citet{janka01}, obtained by assuming zero electron
chemical potential. The right panel of Figure~\ref{f:cooling_heating_app} shows heating by absorption of
electron-type neutrinos as a function of $T_{\nu_e}$, for fixed parameters as shown in the Figure. For reference,
the approximate expression $Q^+_{\nu e} \simeq 160 L_{\nu,52} r_7^{-2} (1-Y_e) (kT_{\nu_e}/4\textrm{MeV})^2$~MeV~s$^{-1}$
\citep{janka01} is also shown, agreeing with our results to within $\sim 10\%$.

\section{Grid of Variable Spacing in FLASH3.2}
\label{s:grid_test}

Time-dependent modelling of post-bounce core-collapse supernova hydrodynamics demands
the ability to resolve a steep density gradient in the protoneutron star atmosphere
($\sim 30$~km) while also allowing for the shock to expand out to at least $\sim 1000$~km
to track an explosion. For a grid in spherical coordinates, an efficient way
to accomplish this is to use variable spacing in radius.

The public version of FLASH3.2 was modified to include this capability, starting
from the existing \emph{uniform grid} mode. The implementation can be
decomposed into two parts. First, defining the cell sizes and coordinates appropriately
when the computational domain is initialized. The second part involves modifying
all the subroutines that assume a uniform grid spacing. 

Grid initialization is accomplished by modifying the \texttt{Grid\_init} 
and \texttt{gr\_create\_domain} subroutines. We define the grid points
in between $r_{\rm min}$ and $r_{\rm max}$ such that consecutive cell
sizes have a ratio $\Delta r_{\rm i+1}/\Delta r_{\rm i} = \zeta > 1$, where $i$
is a cell index which increases with increasing radius.
For a given number of grid cells $N_r$ the ratio can be obtained
by solving (e.g., \citealt{stone92})
\begin{eqnarray}
r_{\rm max}-r_{\rm min} & = & \Delta r_{\rm min}\sum_{i=0}^{N_r-1}\zeta^i\nonumber \\
\label{eq:grid_equation}
                        & = & \Delta r_{\rm min}\left[\frac{\zeta^{N_r} - 1}{\zeta - 1} \right],
\end{eqnarray}
where $\Delta r_{\rm min}$ is the minimum cell size, whose inner edge is located at $r_{\rm min}$.
A logarithmic spacing can be achieved by setting $\zeta = (r_{\rm max}/r_{\rm min})^{1/N_r}$,
which then determines the minimum cell size: $\Delta r_{\rm min} = r_{\rm min}(\zeta-1)$.
It then follows that for $0 \le q \le N_r$,
\begin{equation}
\frac{\Delta r_{q}}{r_q} = \frac{\zeta^q}{(r_q/r_{\rm min})}\frac{\Delta r_{\rm min}}{r_{\rm min}} 
= \frac{\Delta r_{\rm min}}{r_{\rm min}},
\end{equation}
with $r_0 = r_{\rm min}$ and $r_{N_r} = r_{\rm max}$.

There are only three subroutines that make the assumption of a uniform grid when using
spherical coordinates. They are \texttt{hy\_ppm\_sweep}, \texttt{Driver\_computeDt}, and
\texttt{Driver\_verifyInitDt}. In all three cases, a scalar cell spacing was replaced
with a vector in the appropriate locations. The subroutines that compute cell areas
and volumes make direct use of coordinate information, so no additional modification is required
for spherical coordinates.

As a test of the grid implementation, we have run the self-similar explosion problem
of \citet{sedov82}. An energy $E_0$ is initially placed inside some spherical volume with a
radius smaller than some characteristic radius $R_0$. The medium has uniform density
$\rho_0$, and has an ideal gas equation of state with adiabatic index $\gamma$.
Outside of the injection volume, the pressure is uniform an equal to $10^{-5}E_0 R_0^{-3}$.
Our benchmark simulation has $500$ cells uniformly spaced in radius from $r=0$ to $r=R_0$.
The energy is  placed in the first grid cell, with radius $r_{\rm inj}  = 2\times10^{-3}R_0$,
and we set $\gamma=1.4$.

\begin{figure*}
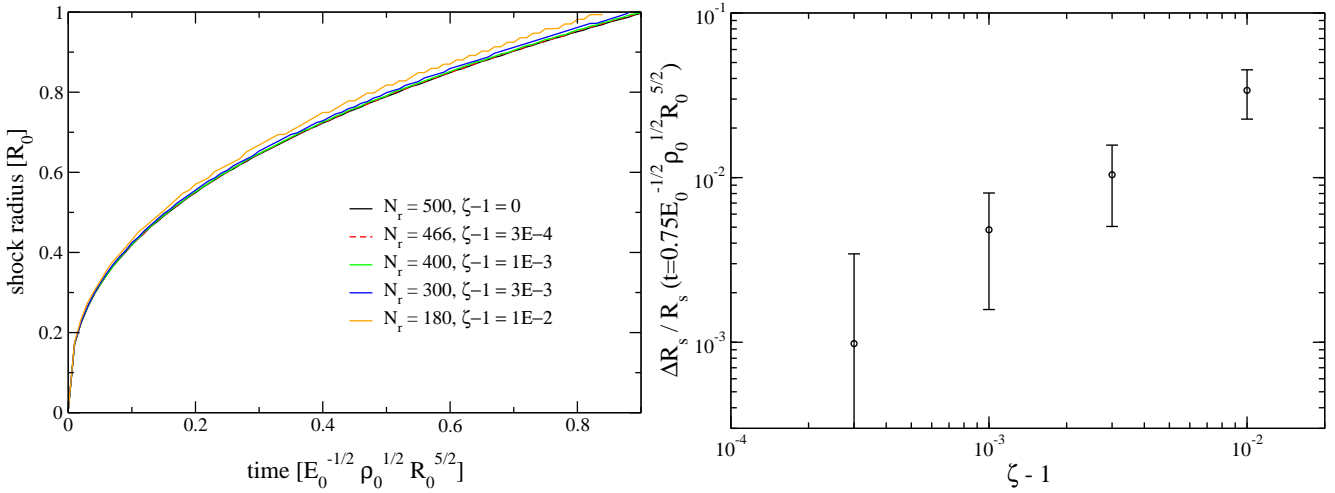

\includegraphics*[width=0.470\textwidth]{f17a.eps}
\includegraphics*[width=0.5\textwidth]{f17b.eps}
\label{f:sedov_test}
\caption{Strong explosion problem of \citet{sedov82}, used here as a test of the non-uniform grid implementation in FLASH3.2.
The left panel shows the shock trajectory as a function of time, with colors labeling
resolutions as shown in the caption. The minimum cell size next to the origin is kept equal to the
uniform grid case $(\zeta=1)$.
The right panel shows the fractional deviation from the
shock position at time $0.75 E_0^{-1/2}\rho_0^{1/2}R_0^{5/2}$ in the $N_r = 500$ run, with error bars indicating the
cell sizes in each simulation at their respective shock positions, normalized to the shock position in the $N_r=500$ run.}
\end{figure*}

We then run a sequence of simulations that preserve the minimum cell spacing 
next to the origin, $\Delta r_{\rm min} = r_{\rm inj}$, but vary the number of cells,
resulting in a ratioed (not logarithmic) grid that satisfies equation~(\ref{eq:grid_equation}). 
The number of cells are $N_r = \{466,400,300,180\}$
which results in $(\zeta-1) \simeq \{3\times 10^{-4},10^{-3},3\times 10^{-3},10^{-2}\}$,
respectively. The left panel of Figure~\ref{f:sedov_test} shows the shock radius (obtained by linear
interpolating the location of the surface with pressure $10^{-3}E_0 R_0^{-3}$) as a function of time for 
the sequence of runs. As the number of cells is decreased relative to the benchmark model, the cells 
in the upper part of the domain
become increasingly coarser, and the shock trajectory gradually diverges from the uniform grid case. 
The right panel of Figure~\ref{f:sedov_test}
shows the fractional deviation from the uniform grid result at a time $0.75 E_0^{-1/2}\rho_0^{1/2}R_0^{5/2}$,
with the error bars indicating the fractional size of the cell at the given shock position, normalized
to the shock position in the benchmark model. As the spacing ratio decreases, deviations tend to zero
close to linearly in $(\zeta - 1)$.

\section{Numerical Calculation of the Rate of Change of Energy}
\label{s:work_integral_calculation}

Here we describe the calculation of the different terms that comprise the rate of change of the total energy
with time (eq.~[\ref{eq:dEdt_terms}]) for Figures~\ref{f:dEdt_oscillatory} and \ref{f:dEdt_nonoscillatory}.

In the Piecewise Parabolic Method \citep{colella84} used in FLASH, the shock is typically broadened along 2-3 cells. 
To compute an accurate shock position, we begin by finding the minimum in the radial derivative of the
pressure. We establish a reference position by taking a weighted average of the radial coordinate, with
weight equal to $|\partial p / \partial r|$ around the minimum. This reference position, which varies smoothly with 
time, is a good tracer of the center of the shock. Because we want quantities below the shock, we define our actual
shock position to be a fixed distance (of the order of two cell widths) behind the center of the shock,
so that the fluid quantities are in the post-shock regime while remaining as close to the shock center as possible.

Next, the energy flux entering through the shock $\dot{E}_{\rm up}$ (eq.~[\ref{eq:Edot_up}]) is computed by interpolating variables
(linearly for velocity and internal energy; logarithmically for pressure and density) to the shock position. The
energy flux leaving the domain $\dot{E}_{\rm dn}$ (eq.~[\ref{eq:Edot_dn}]) is computed at a radius $R_{\rm in}$ 
corresponding to the inner edge
of some cell inside the domain. After some experimentation, we found that the innermost location that is not significantly 
affected by boundary effects is the third active cell from the inner boundary at $R_\nu$. To compute the energy flux at 
$R_{\rm in}$, we average the fluid variables from the cells that this radius separates as an approximation to face-centered 
values. The integrated neutrino source terms $\dot{E}_N$ (eq.~[\ref{eq:Edot_N}])
are computed by simple integration of the cells fully contained between $R_{\rm in}$ and the shock, and then a small differential
term is added by linear interpolation, $4\pi R_{\rm out}^2 \mathscr{L}_{\rm net}\Delta R$, where $R_{\rm out}$ is the outer radius
of the cell that is closest to and fully contained within the shock surface, and $\Delta R = R_S - R_{\rm out}$.
The shock motion term $\dot{E}_S$ is found by first computing the shock velocity through time-centered finite differences, and
then evaluating fluid quantities at the same position as for $\dot{E}_{\rm up}$.

\begin{figure}
\includegraphics*[width=0.5\columnwidth]{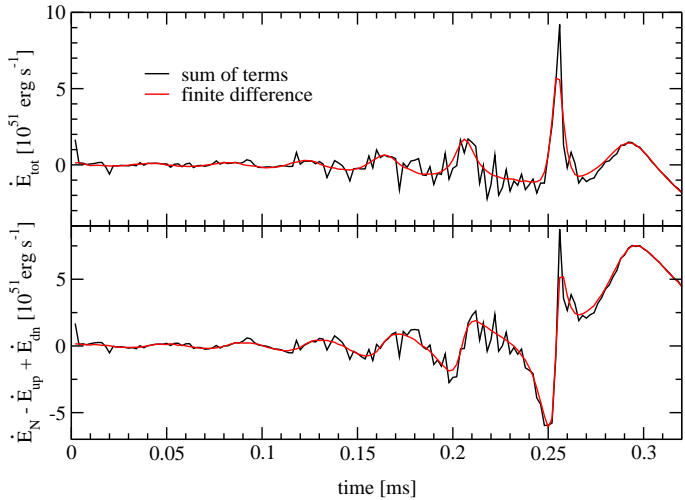}
\caption{\emph{Top:} evolution of the total energy in the model shown in Figure~\ref{f:dEdt_oscillatory}. The black curve
corresponds to the sum of the terms on the right hand side of equation~(\ref{eq:dEdt_terms}), calculated as described
in the text. The red curve is a time-centered finite difference derivative of the instantaneous total energy. 
\emph{Bottom:} evolution of the net energy generation from accretion. The black curve is the sum of the terms
that comprise it, while the red curve is obtained by subtracting the energy generation from shock motion $\dot{E}_s$ from
the red curve in the top panel.}
\label{f:workintegral_test_new}
\end{figure}

For calibration, we also compute the total energy contained between $R_{\rm in}$ and the shock. The total rate of change, computed
through time-centered finite differences, and the sum of the separate terms that make it up (eq.~[\ref{eq:dEdt_terms}])
are shown in Figure~\ref{f:workintegral_test_new}. Quantities are sampled every $2$~ms to eliminate high frequency noise.
Agreement is very good if a zero point, equal to the sum of all the terms that make up $\dot{E}_{\rm tot}$ at $t=0$, is subtracted.

Also shown in Figure~\ref{f:workintegral_test_new} is the net energy generation from accretion, 
$\dot{E}_N - \dot{E}_{\rm up} +\dot{E}_{\rm dn}$,
together with what is obtained if the shock motion term $\dot{E}_s$ is subtracted from the total rate of change of 
the energy $\dot{E}_{\rm tot}$, computed from finite differencing the total energy. Agreement is again very good if the same
zero point is subtracted. This smoother version of the accretion energy generation, together with $\dot{E}_{\rm tot}$
from finite differencing, is what is shown in Figures~\ref{f:dEdt_oscillatory} and \ref{f:dEdt_nonoscillatory}.

\bibliographystyle{apj}
\bibliography{dimension,apj-jour}

\end{document}